# A Survey of Deep Learning Based Software Refactoring

BRIDGET NYIRONGO, Beijing Institute of Technology, China

YANJIE JIANG[*], Peking University, China

HE JIANG, Dalian University of Technology, China

HUI LIU, Beijing Institute of Technology, China

Refactoring is one of the most important activities in software engineering which is used to improve the quality (especially the maintainability) of a software system. The traditional approaches to software refactoring involve designing a series of heuristics for refactoring detection, solution suggestions, and refactoring execution. However, these approaches usually employ manually designed heuristics, which are often tedious, time-consuming, and challenging. With the advancement of deep learning techniques, researchers are attempting to apply deep learning techniques to software refactoring. Consequently, dozens of deep learning-based refactoring approaches have been proposed. However, there is a lack of comprehensive reviews on such works as well as a taxonomy for deep learning-based refactoring. To this end, in this paper, we present a survey on deep learning-based software refactoring. We classify related works into five categories according to the major tasks they cover, i.e., the detection of code smells, the recommendation of refactoring solutions, the end-to-end code transformation as refactoring, quality assurance, and the mining of refactorings. Among these categories, we further present key aspects (i.e., code smell types, refactoring types, training strategies, and evaluation) to give insight into the details of the technologies that have supported refactoring through deep learning. The classification indicates that there is an imbalance in the adoption of deep learning techniques for the process of refactoring. Most of the deep learning techniques have been used for the detection of code smells and the recommendation of refactoring solutions as found in 56.25% and 33.33% of the literature respectively. In contrast, only 6.25% and 4.17% were towards the end-to-end code transformation as refactoring and the mining of refactorings, respectively. Notably, we found no literature representation for the quality assurance for refactoring. We also observe that most of the deep learning techniques have been used to support refactoring processes occurring at the method level whereas classes and variables attracted minimal attention. Finally, we discuss the challenges and limitations associated with the employment of deep learning-based refactorings and present some potential research opportunities for future work.

## 1 INTRODUCTION

Software refactoring is used to improve software quality by changing the internal structure of a system software without altering its external behavior. It is a way of maintaining and improving the quality of software code that helps in the following tasks [26]. First, it helps in the discovery of bugs. When refactoring, more time and work are spent on understanding what the program code does and incorporating any new understanding into the code. This process

[*]Corresponding author

Authors' addresses: Bridget Nyirongo, larjean89@outlook.com, Beijing Institute of Technology, Beijing, China; Yanjie Jiang, yanjiejiang@pku.edu.cn, Peking University, Beijing, China; He Jiang, jianghe@dlut.edu.cn, Dalian University of Technology, Dalian, China; Hui Liu, liuhui08@bit.edu.cn, Beijing Institute of Technology, Beijing, China.

helps to bring to light any assumptions that were previously made, making it less likely that bugs will go unnoticed. Second, refactoring can help to improve the software design. With the development of software, various modifications may be introduced due to misunderstanding of requirements or urgent task assignments. After such modifications, the software will become harder to read and comprehend. Refactoring cleans up the program code as work is done to rearrange and remove parts that are not in order and are unnecessary. This helps to retain the program code's structure thereby improving its design. Third, it helps in rapid software development. Refactoring helps in the faster development of software because it stops the design of the system from going bad. With frequent refactoring, not much time is spent on finding and fixing errors that might arise from poor code design. Fourth, it helps make software easier to understand. This is because a little more time spent on refactoring could make the code better communicate its purpose. This code would say exactly what the developer writing it meant, as such any other developer using this code might easily understand it.

Researchers have dedicated a great amount of time trying to find ways that could make the process of software refactoring less tedious and time-consuming. Different techniques, models, and concepts have been used. Semi-automatic and automatic tools [22, 86, 89, 91, 92, 94, 100] have been developed to assist in the detection, recommendation, and safe application of the refactorings. Most of these tools can easily be integrated as plugins in most of the modern IDEs like Eclipse and Intellij IDE. The integration of these tools in the IDEs is usually alongside the already existing refactoring menus within the IDEs thereby enriching the support rendered towards the process of refactoring. Even though this is the case, it has been noted that most of the approaches used to come up with these tools rely on heuristics which are manual in nature [51, 53]. Thus, researchers have adopted the use of machine learning techniques to minimize the use of manually designed heuristics [23, 25].

Deep learning is a sub-field of machine learning that focuses on creating large neural network models that are capable of making accurate data-driven decisions. Deep learning is mostly suitable for contexts where data is complex and where large datasets are available. The unique aspect of deep learning is the approach it takes to feature design which is characterized by the automatic learning of hierarchical representations from raw data thereby eliminating the need for manual feature engineering. Deep learning models can learn useful features from low-level raw data and complex non-linear mappings from inputs and outputs [29, 43]. This is unlike most of the statistical machine learning models where feature design is a human-intensive task that can require deep domain expertise and consume a lot of time and resources. At the core of deep learning are Artificial Neural Networks (ANN or NN) which are composed of interconnected nodes organized into layers [1, 12, 78]. Deep learning uses several types of Neural Networks each tailored for specific tasks. Some of the most common ones are as follows. *Feedforward Neural Networks(FNN)* also known as Multilayer Perceptrons(MLP). These are mostly used for general-purpose tasks including classification and regression. *Convolutional Neural Networks (CNN)*. CNN utilizes convolutional layers to automatically learn hierarchical features from input images. These were originally designed for image and grid-like data. *Recurrent Neural Networks (RNN)*. RNN contain loops to capture dependencies in sequential data. Two of its variants, Long Short Term Memory (LSTM) and Gated Recurrent Unit (GRU) address the vanishing gradient problem and improve the modeling of long-range dependencies. *Graph Neural Networks (GNN)*. GNN learns to process and extract information from graph-structured inputs. These are used in tasks like node classification, link prediction, and recommendation systems. *Generative Adversarial Networks(GAN)*. GAN consists of a generator network and a discriminator network that are trained simultaneously. These are used for generating new data samples, such as images, text, and music. *Autoencoders*. Autoencoders are neural networks that are used for unsupervised learning and dimensionality reduction. Autoencoders comprise an encoder network to reduce the input data's dimensionality and a decoder network to reconstruct the input data from the reduced representation. *Transformers*. Transformers use a self-attention mechanism to capture relationships between input elements. They are popularised by models like BERT (Bidirectional Encoder Representations from Transformers) and GPT (Generative pretrained Transformer) for NLP tasks, but they have been applied to various other tasks.

Recently, a considerable amount of research [32, 51, 53] has been proposed to investigate and explore the application and adoption of deep learning techniques to automate and support the process of software refactoring. Researchers have employed various deep learning models in the different tasks involved in the process of software refactoring. To present the state-of-the-art on the employment of deep learning in refactoring, researchers conducted literature reviews [2, 56, 63, 110] centering around the use of deep learning for refactoring activities, e.g., the detection of code smells. Naik et al. [63] presented and analyzed 17 related works published from 2016 to 2022 to identify which deep learning techniques had been used for code refactoring and how well they worked. Alabza et al. [2] conducted a systematic review focusing on the deep learning approaches for bad smell detection on 67 studies published until October 2022. Malhotra et al. [56] conducted a systematic literature review examining deep learning's capability to spot code smells on 35 primary studies published from 2013 to July 2023. Zhang et al. [110] conducted a survey on code smell detection based on supervised learning models by analyzing 86 studies published from 2010 to April 2023. Although such reviews have significantly facilitated the understanding of deep learning-based refactoring, we still lack a comprehensive survey that considers the majority of related works, covering all aspects of deep learning-based software refactoring (not just confined to code smell detection), and provides a taxonomy for deep learning-based refactoring.

To this end, in this paper, we conduct a survey by collecting 48 primary studies published from January 2018 to October 2023 and classify them based on the specific refactoring tasks being supported by the deep learning technique, i.e., the detection of code smells, the recommendation of refactoring solutions, the end-to-end code transformation as refactoring, quality assurance, and the mining of refactorings. Under each of these categories, we present key aspects (i.e., code smell types, refactoring types, training strategies) related to the approaches to give insight into the technologies that have supported software refactoring through deep learning. Based on such a presentation, we can provide a more comprehensive perspective on deep learning-based refactoring. To the best of our knowledge, it is the first survey paper that presents the hierarchical taxonomy for deep learning-based software refactoring. In our survey, we attempt to investigate the following research questions:

- **RQ1:** Which tasks in software refactoring have been supported by deep learning techniques, and how often they have been targeted by the surveyed papers?
- **RQ2:** What are the common deep learning techniques used in software refactoring?
- **RQ3:** How effective is the use of deep learning models in the process of software refactoring?
- **RQ4:** What are the limitations and challenges associated with the use of deep learning techniques in software refactoring?

Our classification indicates that there is an imbalance in the refactoring tasks which have been supported by deep learning techniques. Our survey indicates that most of it has been towards the detection of code smells and the recommendation of refactoring solutions, unlike the end-to-end code transformation as refactoring, quality assurance, and mining of refactorings. Thus, there is a need for more future work to address the current imbalance, specifically in areas of deep learning supporting the application, quality assurance, and mining of refactorings. The rest of the paper is structured as follows: Section 2 introduces related work. Section 3 presents the methodology used for the survey. Section 4 presents and discusses studies on the detection of code smells. Section 5 presents and discusses studies on the recommendation of refactoring solutions. Section 6 presents and discusses studies on end-to-end code transformation as refactoring. Section 7 presents and discusses studies on the mining of refactorings. Section 8 discusses some of the challenges and opportunities associated with deep learning-based refactoring, and section 9 concludes the survey.

## 2 RELATED WORK

Naik et al. [63] conducted a systematic review of the current studies on deep learning-based code refactoring. The survey presented a high-level analysis of 17 primary works published from 2016 to 2022. The key insight of this survey was to present the state-of-the-art in deep learning-based code refactoring. The review addressed research questions whose main focus was on the commonly used deep learning techniques and the performance of the deep learning-based refactoring approaches. This review indicated that CNN, RNN, and GNN are the commonly used deep learning models for code refactoring with Multilayer Percepton (MLP) performing the best. They also noted that most of the existing studies focus on Java code, method-level refactoring, and single-language refactoring with various evaluation methods. Compared to the survey by Naik et al. [63], our survey covers substantially more related works, increasing the number of surveyed papers from 17 to 48. Our well-designed search strategy retrieved many closely related papers missed by their survey. Another difference is that we present a comprehensive and hierarchical taxonomy of deep learning-based refactoring, and classify all related works based on the taxonomy.

Alabza et al. [2] conducted a review on deep learning-based approaches for bad smell detection, and their focus was to summarise and synthesize the studies that used deep learning for bad smell detection. They collected and analyzed 67 studies until October 2022. They analyzed deep learning models concerning the purpose of the model, the detected bad smells, the employed training datasets, features, pre-processing techniques, and encoding techniques used for feature transformation. Notably, this survey involved all kinds of code smells. This review indicated that code clones are the most recurring smell. The review showed that supervised learning is the most adopted learning approach used for deep learning-based code smell detection. Also, they observed that CNN, RNN, DNN, LSTM, Attention models, and Autoencoders are the most popularly used deep learning models. Notably, this review focused only on the use of deep learning models for the detection of code smells which is one of the tasks used for the identification of refactoring opportunities. In contrast, our survey covers all aspects of refactoring.

Malhotra et al. [56] conducted a literature review examining deep learning's capability to spot code smells. They presented a total of 35 primary study works from the years 2013 to 2023. The consolidated studies highlighted four key concepts, that is, the types of code smells addressed, the deep learning approach utilized in the experiment, evaluation strategies employed in the studies, and the performance analysis of the model proposed. This review showed that the most common code smells detected include feature envy, god class, long method, complex class, and large class. It also indicated that the most common deep learning algorithms used are RNN and CNN, often combined with other techniques for better results. Notably, this systematic analysis did not focus on the whole refactoring process (as what we do). Instead, they only focused on the recommendation of refactoring solutions and the end-to-end code transformation as refactoring.

Zhang et al. [110] conducted a survey on code smell detection based on supervised learning models. They surveyed 86 papers from January 2010 to April 2023. They formulated a total of 7 research questions which were empirically evaluated from different aspects such as dataset construction, data preprocessing, feature selection, and model training. Based on their analysis they concluded that most of the existing works suffer from issues such as sample imbalance, different attention to types of code smell, and limited feature selection. They also made suggestions for future work, one of which involves exploring the correlation between features and the perspective of code smells within the context of model interpretability. Notably, the core focus of this paper was on the types of deep learning models used for the detection of code smells, and not the use of deep learning models in the process and support of software refactoring. Our paper differs from such reviews in that they focus on code smell detection only whereas we cover a much larger scope, i.e., the whole process of software refactoring.

## 3 METHODOLOGY

We surveyed deep learning-based software refactoring by exploring and searching the following databases: https://dl.acm.org, http://ieeexplore.org, https://springer.com, https://sciencedirect.com, https://onlinelibrary.wiley.com/, and https://scholar.google.com. We used these databases since they contain a comprehensive coverage of academic publications, conferences, and journals in the field of software engineering. Utilizing these databases ensured a thorough and inclusive exploration of the existing literature, enabling a comprehensive understanding of the landscape of deep learning-based refactoring techniques and advancements. We used a set of keywords based on the research questions outlined in Section 1 to retrieve studies from these databases. The main keywords were "deep learning", "software refactoring", and "code smells". We also included synonyms of these keywords, such as "refactoring", "code refactoring", and "bad smells". These search terms were used with advanced search filters within the databases(e.g., specific year range (2018 to 2023) and language specification (English)). For example, on ACM, we used the following search query: *"query": (("deep learning") AND ("refactoring")), "filter": publication date: 01/01/2018 TO 10/31/2023, owners.owner=HOSTED*. We obtained a total of 1,755 papers after the filtering, and selected the top 100 (if there are more than 100) from each database based on relevance, resulting in 486 papers. The selection of studies was limited to the top 100 papers from each database based on relevance because our initial analysis suggests that items outside the top 100 are often outside the scope of the survey.

The first author conducted a manual search to decide whether each paper should be included or not. This was done by reading through the title and abstract of each candidate paper. To ensure the correctness of the manual search, inclusion criteria were defined, and continuous and open communication was maintained among the authors. The main criterion for the classification was focused on papers that used deep learning for refactoring tasks, such as detecting code smells, recommending refactorings, and the end-to-end code transformation as refactoring. The use of this criterion resulted in remaining with a total of 35 papers. To enhance the comprehensiveness of our literature search, we conducted a snowballing process by scrutinizing the reference sections of the initial set of papers. This iterative process involved exploring citations within these papers to identify additional relevant studies. Google Scholar was utilized as the primary database for this snowballing process, ensuring an expansive and thorough exploration. As a result of this snowballing procedure, we identified and included 13 more papers that were deemed pertinent to our survey focus. Combining these newly discovered papers with the initial set, a total of 48 papers were meticulously collected and considered as primary studies for our survey. This approach allowed us to cast a wider net in the literature search, ensuring the inclusion of studies that may not have been initially captured, thereby enriching the depth and scope of our survey. Figure 1 presents the papers on deep learning-based refactoring from January 2018 to October 2023. The figure shows that there is a growing interest among researchers to adopt deep learning techniques for software refactoring.

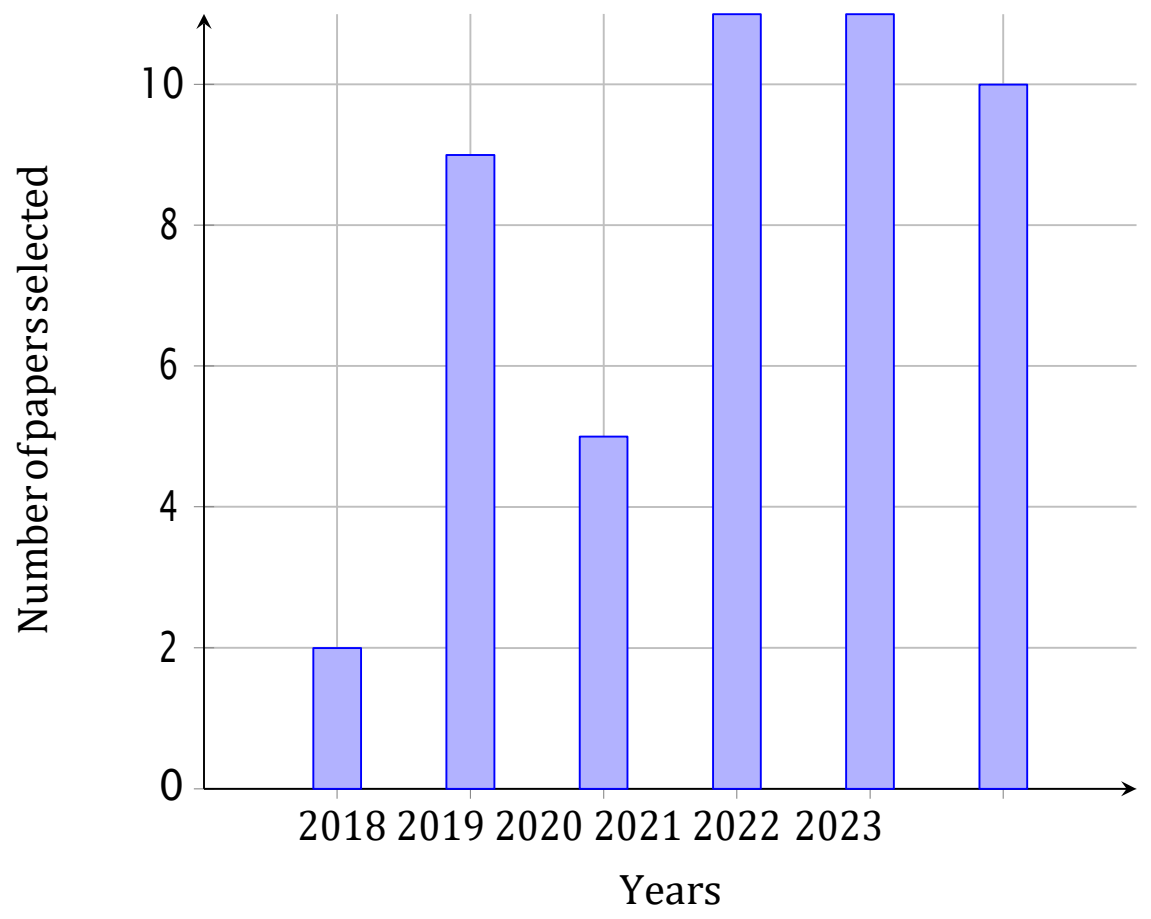

Fig. 1. Primary studies through the years

From the collected data we have developed a taxonomy classification hierarchy as presented in Figure 2. The classification hierarchy for the taxonomy is based on the software refactoring tasks that could be supported by deep learning techniques. These tasks include the detection of code smells, recommendation of refactoring solutions, end-to-end code transformation as refactoring, quality assurance, and the mining of refactorings. Figure 2 presents the total number of papers collected for each specific task (as presented on the lower right corners of the nodes). Our survey indicates that there is an imbalance in the adoption of deep learning techniques for the process of refactoring. Most of the deep learning techniques have been used to detect code smells and recommend refactoring solutions as found in 56.25% and 33.33% of the literature respectively. In contrast, only 6.25% and 4.17% were towards the end-to-end code transformation as refactoring and the mining of refactorings, respectively. Notably, we found no literature representation for the quality assurance for refactoring. We have also observed that some researchers made contributions to more than one refactoring task in the use of deep learning. For instance, 22.45% of the studies focused on applying deep learning techniques on more than one refactoring task, i.e., the detection of code smells and recommendations for refactoring solutions. For such papers, we discuss them in different sections regarding their contribution to the different tasks. As suggested by the legend in the left bottom of Figure 2, the leftmost node presents the root of the taxonomy, and the nodes in the following layer represent different steps involved in software refactoring. Nodes in the rightmost two layers represent different aspects (factors) that could be employed to further classify related works within the given category. For example, the "*code smell types*" node on the third layer suggests that we may classify related works on "*detection of code smells*" into sub-categories according to the code smells' types they support.

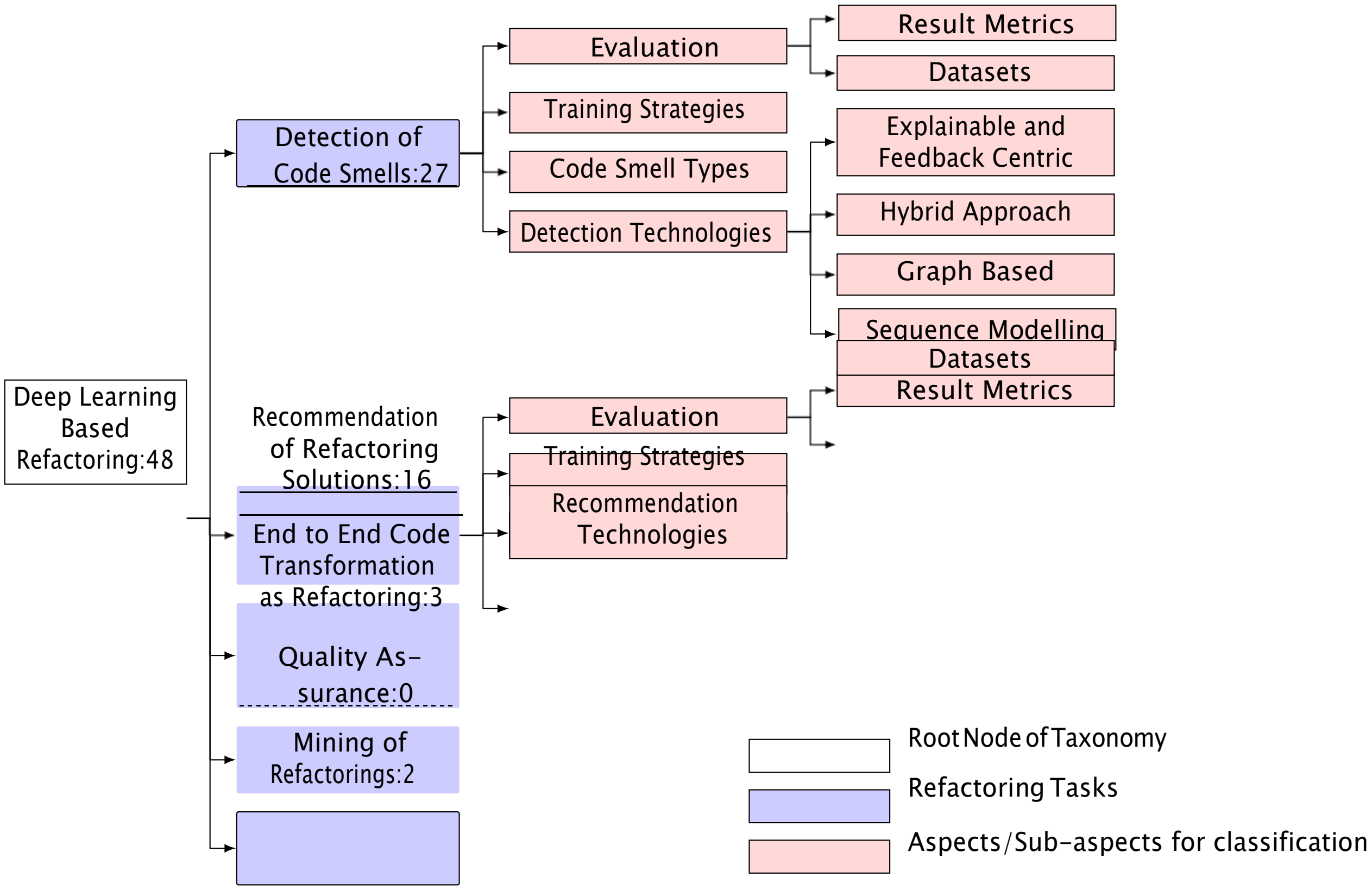

Fig. 2. Taxonomy of deep learning-based refactoring

## 4 DETECTION OF CODE SMELLS

Refactoring is a crucial task in software engineering. To conduct refactoring, there needs to be awareness of the issues in the code that might call for the process of refactoring. One of the most common issues that trigger the need for refactoring is code smells. Code smells are certain structures in the code that indicate the possibility of refactoring [26]. Detecting code smells manually on a large codebase is a challenging task. Therefore, researchers have explored different techniques, including automatic and semi-automatic ones, to aid in the detection of code smells. Despite various techniques proposed by researchers, most of them rely on manually designed heuristics. The manual heuristics make the process of detecting code smells for refactoring opportunities time-consuming. To solve this issue, researchers have started exploring deep-learning techniques for the detection of code smells. Thus, researchers have proposed various deep-learning techniques for the detection of code smells. In this section, we discuss and present these deep learning techniques based on the detection technologies, code smell types, training strategies, and evaluation techniques employed.

### 4.1 Detection Technologies

Researchers [51, 53, 80, 109] have used various deep-learning approaches for the detection of code smells. The details of these approaches are presented as follows:

*4.1.1 Sequence Modelling Based Approaches.* This category pertains to technological approaches that focus on the sequence of code, such as source code tokens or characters. The approaches in this group utilize deep learning techniques to capture contextual dependencies within code snippets. Also, these approaches detect code smells based on the sequential nature of code. CNN, RNN, LSTM, Transformers, etc. are some of the deep learning techniques employed in this category. The deep learning techniques in this category can effectively learn patterns and relationships within code sequences, thus aiding in the process of code smell detection.

Liu et al. [53] proposed a deep learning-based approach to detecting feature envy, which is one of the most common code smells. They used a Convolutional Neural Network (CNN) as their deep neural network-based classifier. The classifier's input was divided into two parts: textual input and numerical input. The textual input was a word sequence that consisted of the method's name, the name of its enclosing class, and the name of the potential target class. The information in the textual input had to pass through an embedding layer that converted the text description into numerical vectors. The numerical vectors were then fed into the CNN. They had three CNN layers, each with filters=128, kernel size=1, and activation Tanh. The CNN classifier was trained automatically, without any human intervention. The labeled samples for training were generated automatically based on open-source applications. To evaluate the approach, they conducted a two-part verification process. First, they evaluated the approach on 7 well-known open-source applications (Junit, PMD, JExcelAPI, Areca, Freeplane, jEdit, and Weka) without automatically injecting feature envy smells. This first evaluation gave an improvement on F-measure against state-of-the-art by 34.32%. The second part of the evaluation was carried out on 3 open-source applications (XMD, JSmooth, and Neuroph) where no smells were injected, and this outperformed the state-of-the-art.

Liu et al. [21] proposed a method called feTruth to enhance the detection of feature envy in software using deep learning with real-world examples. The feTruth technique used evolutionary histories of open-source projects stored in version control systems such as GitHub to extract real examples of feature envy. The extracted real-world examples were then used to train a deep learning-based prediction model. During the testing phase, feTruth would examine the source code of a software project and generate a list of feature envy occurrences associated with methods in the project. feTruth included a heuristics-based filter and a learning-based filter. These two filters were used to exclude false positives reported by RefactoringMiner [93]. The heuristic-based filter would exclude false positives if the source class of the potential refactoring did not exist in the new version or if the target class of the potential refactoring did not exist in the old version. The learning-based filter leveraged a decision tree-based classifier to distinguish false positives from true positives based on a sequence of features of the refactorings. By using these techniques, the researchers were able to generate high-quality and large-scale training data for feature envy detection. They used a CNN in the design for the deep learning-based classifier. The CNN classifier leveraged new features not yet exploited by existing approaches. The feTruth method was compared against Liu's approach [51], JDeodorant [22], and JMove [89]. The subjects for their projects were divided into two parts. The first part consisted of 500 Java projects, which were used to discover real-world examples of feature envy. These projects were collected from GitHub by selecting the top 500 most popular projects with the largest number of stars. The second part consisted of 5 open-source Java projects, which were used to evaluate the proposed approach and the selected baseline. These 5 projects were chosen from Defects4J [42]. The evaluation results on real-world open-source projects suggested that the proposed approach substantially outperforms the state-of-the-art in the detection of feature envy smells. The approach improves the precision and recall in feature envy detection by 38.5%.

Das et al. [18] proposed a deep learning approach to detect brain class and brain method code smells in software applications. They used Convolutional Neural Networks (CNN) to train a neural network-based classifier. The approach was based on a large corpus of software applications, which generated a huge number of training samples. These samples were labeled to indicate whether they were a code smell of kind brain class and brain method. The neural network used in this approach had several layers - the first layer had a one-dimensional CNN layer with 256 filters and a kernel size of 1. The activation function used was Tanh. The second layer was also a one-dimensional CNN layer with 128 filters and Tanh as the activation function. They added a flattened layer as the third layer to connect the convolution layer with dense layers. The fourth layer had a dense layer with 128 filters and ReLU as the activation function. The last layer was a dense layer with only one filter and Sigmoid as the activation function. This layer acted as the output layer. They used 30 open-source Java Projects as subject applications, acquired through sharing activities in GitHub repositories. The dataset of the Java projects was split into mutually exclusive training and test sets. The experiment demonstrated high-accuracy results for both code smells.

Lin et al. [48] proposed a new approach for detecting code smells. They used a full convolutional network that could identify and use local correspondences by making use of semantic features. They defined a multidimensional array, *h*w*d*, to represent the convolutional network (*where h and d are space dimensions and d is the channel*). For their experiment, they used an open-source database that was initially published by the author. They used this database to detect various code smells such as long method, lazy class, speculative generality, refused bequest, duplicated code, contrived complexity, shotgun surgery, and uncontrolled side effects.

Liu et al. [53] proposed an approach for code smell detection using deep learning. However, this approach had some limitations. To address these limitations, they presented a new approach in [51]. This approach was generic and evaluated on four code smells: feature envy, long method, god class, and misplaced class. They improved the deep neural network used in [53] by using bootstrap aggregating. They used a classifier that generated several bootstrap samples simultaneously from a given training dataset. The classifier trained multiple binary classifiers that in turn determined the final classification by voting. The study highlighted that it is difficult to design and train a generic classifier to detect all code smells since different features are needed for different smells. Therefore, they presented different classifiers for different code smells. They used a Convolutional Neural Network (CNN) for the detection of feature envy and misplaced class. A dense layer-based classifier was used for the detection of the long method. Dense layers coupled with Long Short Term Memory(LSTM) were used for the detection of the god class. The *classifier used for feature envy* was similar to the one in their earlier work [53]. The deep neural *classifier used for the long method* was composed of five dense layers besides the input and output layers. The resulting features which were extracted by the hidden layers were fed into the output layer. This process mapped the features into a single output to suggest if the feature is associated with long method smell. The *classifier for the god class* was composed of two parts: a textual part and a numerical (code metric) input. The textual input was a word sequence formulated by concatenating the names of attributes and methods declared within the class under test. After converting the textual input into numerical vectors through embedding, the resulting vectors were handled by an LSTM layer. This was unlike with the code metrics which were fed straight into a dense layer. This output was then merged with the output of the LSTM before being fed into another dense layer. The output of the dense layer was the one that indicated whether a class should be decomposed or not. The *classifier employed for the detection of the misplaced class* was similar to that of feature envy because both these smells are caused by misplaced software entities like methods, classes, etc. They evaluated the approach on 10 open-source applications. The approach was compared against JDeodorant [22], DÉCOR [61], and TACO [68]. The results indicated that the approach outperformed the state-of-the-art. They also evaluated the proposed approach on real-world applications without any injection of smells. The two-step evaluation of the approach using generated data and real-world data gave considerable differences in the performance of their proposed approach. This led to the conclusion that perhaps the evaluation of code smell detection approaches should rely more on manually validated testing data that are often more reliable than generated data.

*4.1.2 Graph Based Approaches.* This category refers to the use of deep learning techniques to analyze the structural properties of code representation. Specifically, models like GNN and GCN are utilized to detect code smells by analyzing the structural relationship between different elements in the code. By using these models, complex dependencies and interactions can be captured, thereby improving the detection of code smells.

Yu et al. [105] proposed a Graph Neural Network (GNN) based approach to address the issue of inherent calling relationships between methods that often result in low detection efficiency for feature envy detection. To achieve this approach, the authors collected code metrics and calling relationships. The collected features were then converted into a graph where nodes represented the code metrics of a method and edges represented the calling relationships between methods. To address the imbalance of positive and negative samples, they introduced a graph augmenter to obtain an enhanced graph. They then fed the enhanced graph into a GNN model for training and prediction. The GNN classifier had four layers: the input layer, the GraphSAGE layer, the dropout layer, a fully connected layer, and an output layer. The input layer received the augmented graph after oversampling. The GraphSAGE updated the embedding of nodes based on the embedding space. The dropout layer prevented the classifier from overfitting during node classification training. The fully connected layers converted the output of the dropout layer into a one-dimensional vector for final classification. The output layer received the vector and outputted the prediction of the classifier through the activation function Sigmoid. For their experiment, the authors used five open-source projects (BinNavi, ActiveMQ, Kafka, Alluxio, and Realm-java) collected from a dataset labeled by Sharma and Kessentini [82]. The dataset contained 86,652 open-source projects mainly written in Java and C# on GitHub. To select the five projects, they considered projects whose updates were: 1) within two years, 2) had more than 2,000 stars on GitHub, and 3) had more than 5,000 methods and 500 classes. The approach for detecting the feature envy smell achieved an average F1-score of 78.90%, which is 37.98% higher than other comparison approaches.

Hanyu et al. [35] proposed a graph-based deep learning approach to detect long methods. Their approach extended the Program Dependency Graph (PDG) into a Directed-Heterogeneous Graph. The Directed-Heterogeneous Graph was then used as the input graph. They employed the Graph Convolutional Network (GCN) to construct a graph neural network for long method detection. The input in this approach consisted of two kinds of nodes (method node and statement node) and four types of edges (include edge, control flow edge, control dependency edge, and data dependency edge). The Graph Convolutional Network had two layers and one linear layer. To obtain enough data samples for the deep learning classifier, they introduced a semi-automatic approach to generate a large number of data samples. To validate their approach, they compared it with existing methods using five groups of manually reviewed datasets.

*4.1.3 Hybrid Approaches.* This category refers to the use of deep learning techniques in combination with other methods to enhance the accuracy and effectiveness of code smell detection. For instance, deep learning techniques such as CNN, GNN, and attention mechanisms can be combined to leverage the strengths of each approach. Hybrid-based approaches aim to gain a more comprehensive understanding of the code, potentially leading to better performance in detecting code smells.

*Combination of Structural and Semantic Features.* Zhang et al. [109] proposed a new approach called DeleSmell to detect code smells using a deep learning model and Latent Semantic Analysis (LSA). They argued that most of the existing approaches suffer from two things: 1) incomplete feature extraction and 2) an unbalanced distribution between positive and negative samples. To address these issues, they developed a refactoring tool to transform good source code into smelly code and generate positive samples based on real-world project cases. They built a dataset with over 200,000 samples from 24 real-world projects to improve dataset imbalance. DeleSmell collected both structural features through iPlasma and semantic features via Latent Semantic Analysis and Word2Vec. DeleSmell's model comprised a CNN branch, a Gate Recurrent Unit (GRU)-attention branch, dense layers, and an SVM branch. The input was processed by the GRU-attention branch and CNN branch in parallel. The CNN branch had a feature extraction component followed by a classification component. The feature extraction component included a set of hidden layers, including convolution, batch normalization, and dropout layers. The output layer of the last dropout layer was connected to the input of a densely connected network that comprised a stack of two dense layers. An attention mechanism was introduced in the GRU branch to learn the important features in the dataset while suppressing the interference of irrelevant information on the classification results. For the SVM, the kernel method was used to map the nonlinear samples to high dimensional space and identify the optimal hyperplane by maximizing the classification interval between the two samples. The input of the last dense layer consisted of features concatenated by the GRU attention branch and the CNN branch and was connected to the SVM for the final classification. Grid search was used to tune the hyperparameters of the classifiers. ReLU was used as the activation function of this approach. DeleSmell was used to detect brain class and brain method code smells.

In their research, Ma et al. [55] explored the use of a pre-trained model called CodeT5 to detect feature envy, one of the most common code smells. They also investigated the performance of different pre-trained models on feature envy detection by comparing CodeT5 with two other models, CodeBERT and CodeGPT. CodeT5 is an encoder-decoder model that considers the token type information in code. CodeGPT is a transformer-based language model that is pre-trained on programming languages for code completion and text-to-code generation tasks. CodeBERT is a multilayer transformer model that uses the same JavaTokenizer as CodeT5 to extract token sequences from source code. The researchers used these models to extract semantic relationships between code snippets and compared their performance. They evaluated their approach on ten open-source projects (Junit, PMD, JExtractAPI, Areca, Freeplane, JEdit, Weka, Adbextract, Aoi, and Grinder). Liu et al.'s [51] approach was used as their baseline for comparison. The results showed that their approach improved the F-measure by 29.32% on feature envy detection compared to the state-of-the-art.

In their study, Hadj-Kacem and Bouassida proposed a hybrid approach for detecting code smells using deep autoencoder and Artificial Neural Network (ANN) [32]. Both unsupervised and supervised algorithms were used to identify the code smells. The approach had two phases. In the first phase, a deep autoencoder was used for dimensionality reduction, which extracted the most relevant features. Once the feature space was reduced with a small reconstruction error, the ANN classifier would then learn the newly generated data and output the final results. The second phase used a supervised learning classification by using the ANN. The approach was applied to four code smells: god class, data class, feature envy, and long method. The study adopted a set of four datasets that were extracted from 74 open-source systems. The results showed high accuracy with precision and recall values. The best F-measure value was 98.93%, which was achieved with the god class code smell. Even at the method level, the F-measure surpassed 96%. These results validated the effectiveness of the approach.

Sharma et al. [80, 81] conducted a study on the feasibility of using deep learning models for detecting code smells without the need for extensive feature engineering. They investigated the possibility of applying transfer learning in this context. The researchers trained smell detection models based on Convolutional Neural Networks (CNN), Recurrent Neural Networks (RNN), and autoencoder models. The *CNN* layer consisted of a feature extraction part followed by a classification part. The feature extraction part was composed of an ensemble of layers, including convolution, batch normalization, and max pooling layers. These layers formed the hidden layers of their architecture. The convolution layer performed convolution operations based on the specified filter and kernel parameters. The convolution layer also computed the network's weights to the next layer. The max pooling layer reduced the dimensionality of the feature space. The batch normalization layer mitigated the effects of varied input distribution for each training mini-batch which optimized the training. The output of the max pooling layer was connected to the dropout layer, which performed another regularization by ignoring some randomly selected nodes during training to prevent overfitting. The output of the last dropout layer was fed into a densely connected classifier network that had a stack of two dense layers. These classifiers processed one-dimensional vectors, whereas the incoming output from the last hidden layer was a three-dimensional tensor. For this reason, the flattened layer was used first to transform the data into the appropriate format before feeding them into the first dense layer with 32 units and ReLU activation. This was followed by the second dense layer with one unit and Sigmoid activation. This second layer comprised the output layer and contained a single neuron to make predictions on whether a given instance belongs to the positive or negative class in terms of smell investigation. The layer used the Sigmoid function to produce a probability within a range of 0 to 1. The *RNN* was comprised of an embedding layer followed by a feature learning part (a hidden LSTM layer). It was succeeded by a regularization (a dropout layer) and classification (a dense layer) part. The embedding layer mapped discrete tokens into compact vector representations. To avoid the noise produced by the padded zeros in the input arrays, they set the mask zero parameters by the Keras embedding layer implementation. Thus, the padding was ignored, and only meaningful parts of the input data were taken into account. The dropout and the recurrent dropout parameters of LSTM were set to layer 0.1. The output from the embedding layer was fed into the LSTM layer, which, in turn, gave output to the dropout layer. The training and evaluation samples for this approach were generated by downloading repositories containing C# and Java code from GitHub after filtering out low-quality repositories by RepoReapers. They downloaded 922 C# and 922 Java repositories in total. They applied this technique for detecting complex method, complex conditional, feature envy, and multifaceted abstraction. Through this study, they discovered that although deep learning methods could be used for code smell detection, the performance is smell-specific. That is, it is very difficult to have a simple and direct solution for this. They also noted that they could not find a clear superior method between one-dimensional and two-dimensional CNNs. One-dimensional CNNs performed slightly better for the smells 'empty catch block and multifaceted abstraction', while two-dimensional CNNs performed better than their one-dimensional counterpart for 'complex method and magic number' [80].

In their paper, Hadj-Kacem and Bouassida proposed a method for detecting code smells in software using a deep learning algorithm [33]. They used an abstract syntax tree and a variational autoencoder to extract semantic information from the source code. Firstly, they parsed the source code into the AST and transformed each tree into a vector representation that was then fed into the variational autoencoder. The autoencoder generated a latent representation which was used to reconstruct the original input data. A logistic regression classifier was then applied to determine whether the code was a code smell or not. The approach was evaluated on the Landfill dataset [67]. The results showed that the proposed method was effective in detecting code smells such as blob, feature envy, and long method.

Xu and Zhang [102] proposed a deep-learning approach to detect code smells based on Abstract Syntax Trees (ASTs). The approach captures the structural and semantic features of code fragments from the ASTs, by utilizing sequences of statement trees. The sequences of statement trees were encoded using bi-directional GRU and maximum pooling. Then, semantic and structural features were extracted from the encoded sequence to obtain final vector representations of the code fragments. The approach was applied to four types of code smells: insufficient modularization, deficient encapsulation, feature envy, and empty catch block. The final detection results were obtained through fully connected layers. The approach was applied to 500 high-quality Java projects from GitHub, outperforming state-of-the-art deep learning models for both small-grained and larger-grained code smells.

*Attention Mechanism and Enhanced Neural Networks*. Zhang and Dong proposed a new approach for detecting code smells called MARS, which is based on a Metric-Attention-based Residual network [108]. This approach was used to identify brain class and brain method code smells. MARS addresses the issue of gradient degradation by utilizing an improved Residual Network (ResNet). The reason they chose ResNet is that it can reduce model parameters while speeding up the training process. ResNet extracts deep feature information to enhance the accuracy of code smell detection. The approach increases the weight value of important code metrics to label smelly samples by introducing a metric attention mechanism. The attention mechanism used in this approach was inspired by SENet [37]. The approach comprised a fully connected layer, used Tanh as the activation function to accelerate the convergence speed of the model, and used Sigmoid as the gated function. The improved ResNet had a convolution layer, a batch normalization layer that accelerated the convergence speed of the network, used ReLU as the activation function, and had an addition as the sum operation. To train MARS, they extracted more than 270,000 samples from 20 real-world applications to generate a dataset called BrainCode, which is publicly available. They evaluated the effectiveness of the proposed approach by answering five research questions. The results showed that MARS achieved an average of 2.01% higher accuracy than the existing approaches.

Zhao et al. [111] developed a model to detect feature envy, which is based on dual attention and correlation feature mining. Firstly, they proposed a strategy for entity representation using multiple views. This strategy increased the model's robustness and improved the correlation feature and the model's suitability. Secondly, they added an attention mechanism to CNN's channel and spatial dimensions. The addition of the attention mechanism enabled the accurate capturing of the correlation features between entities and controlled the information flow. They compared their approach against Liu's [53] method, JMove [89] and JDeodorant [22]. Five open-source projects were used as subject applications. The experimental evaluation was divided into two parts: 1) large-scale data with feature envy automatically injected for training and classifier verification, and 2) small-scale data without feature envy injected for evaluating the approach's effectiveness on real projects. The evaluation results for both feature envy-injected and non-injected projects showed that the proposed approach outperformed the state-of-the-art.

In their paper, Wang et al. proposed a new model for detecting feature envy using a Bi-LSTM with self-attention mechanism [98]. They approached the problem as a deep learning task and used two input parts: a simpler distance metric and text features extracted from the source code. Their approach consisted of three modules: text feature extract for processing the text input, distance value enhancement for handling the distance metric input, and a feedforward neural network for classification. Notably, they introduced a basic attention function (AttentionBasic) that is widely used in language modeling, in addition to the three score functions related to attention mechanism (AttentionAdd, AttentionDot, and AttentionMinus) [98]. They evaluated their approach using a dataset generated from seven open-source Java projects, which was released by Liu et al. [51]. The results showed that their approach outperformed JDeodorant [22], JMove [89], and the first deep learning method proposed by Liu et al. [51].

Guo et al. [31] proposed a method to detect feature envy code smell using a deep semantics-based approach that combined method representation and a CNN model. The method representation technique was based on an attention mechanism and an LSTM network. The LSTM network represented the textual information of methods in source code. This technique extracted semantic features from textual information and reflected the contextual relationships among code parts. The attention mechanism helped in extracting specific features that were significant to code smell detection. The semantic features supplemented the limited structural information in the code metrics and improved the accuracy of code smell detection. To achieve this approach, they first converted the textual information into vectors representing the description of the method using the method representation. In the second part, the metric cluster was fed into a CNN-based model which had three convolutional layers and did not set the pooling layer. The CNN could fully extract the features from the structural information in the code metrics to reflect the relations between adjacent metrics. After this, they applied a flattened layer to turn the shape of the input into a one-dimensional layer. In the third part, they used a Multilayer Perceptron-based neural network. The outputs of the CNN model and method representation were connected at the connection layer, which concatenated all inputs including the features from the textual input and the metrics input. Behind the connection layer, they set two dense layers and one output layer to facilitate the final classification which mapped the textual input and the metrics input into a single output. The output layer had only one neuron which represented the result of the identifier, i.e., smelly or non-smelly. Sigmoid was used as the activation function. The approach was evaluated using a dataset from 74 open-source projects. The results suggested that the approach achieved significantly better performance than the state-of-the-art approaches.

In a study by Liu et al. [54], an automated method was proposed to spot and refactor inconsistent method names. They used a graph vector and Convolutional Neural Network (CNN) to extract deep representations of the method names and bodies, respectively. The approach worked by computing two sets of similar names when given a method name. The first set included those that could be identified by the trained model of method names. The second set included names of methods whose bodies were positively identified as similar to the body of the input method. If the two sets intersected to some extent, the method name was identified to be consistent. If not, it was identified as inconsistent. They then leveraged the second set of consistent names to suggest new names when the input method was flagged as inconsistent. The CNN used in this approach had two pairs of convolutional and subsampling layers. These layers were used to capture the local features of methods and decrease the dimensions of input data. The network layers from the second subsampling layer to the subsequent layers were fully connected. This meant that they could combine all local features captured by convolutional and subsampling layers. For this approach, the output of dense layers was chosen to be the vector representation of method bodies, which synthesized all local features captured by the other layers. The researchers collected both the training and test data from open-source projects from four different communities (Apache, Spring, Hibernate, and Google). They only considered 430 Java projects with at least 100 commits to ensure that the projects had been well maintained. The experimental results showed that the approach achieved an F-measure of 67.9% on identifying inconsistent method names, improving about 15 percentage points over the state-of-the-art.

Li and Zhang [47] proposed a hybrid model with a multi-level code representation to optimize code smell detection. They first parsed the code into an Abstract Syntax Tree (AST) with control and data flow edges. Then, they applied a Graph Convolutional Network to get the prediction at the syntactic and semantic level. Next, they analyzed the code token at the token level using the bidirectional Long Short Term Memory network with an attention mechanism. They applied this approach to a total of 9 code smells (magic number, long identifier, long statement, missing default, complex method, long parameter list, complex conditional, long method, empty catch clause, and multi smells) and combined them to come up with a multi-labeled dataset.

Zhang and Jia [107] proposed a new technique to detect feature envy using self-attention and Long Short Time Memory (LSTM). They were inspired by the transformer model in the formulation of this approach. The researchers added positional encoding to preserve the meaning of sequences and compensate for the lack of positional information in the pure attention mechanism. They built on the existing deep learning model proposed by Liu et al. [51]. Also, the researchers considered attention mechanism, LSTM structure, and snapshot ensemble to improve the detection performance. In their approach, they utilized self-attention as the attention mechanism. This mechanism was implemented with the positional encoding layer right after the embedding layer. The self-attention score was calculated using the Softmax function and then multiplied with the original input to obtain the word embedding vector with attention score. They also included an LSTM block after the attention layer to extract deeper semantic information. For the CNN model, they initially tried adding a pooling layer but found it to be less effective than a dense layer. They changed the convolution block's structure from 128 dimensions to 64 dimensions and then to 32 dimensions. The researchers believed that this structure could filter out the critical features while reducing the consumption of time and computing time. After the CNN, they inserted a dense layer with 1024 nodes right after concatenation. They also included the concept of integrated learning by incorporating a snapshot ensemble in their approach, inspired by Huang et al. [39]. However, they proposed a new periodic function instead of using the cosine function to adjust the learning rate. The results showed that their model achieved better performance on four evaluation metrics, with precision increasing by 0.048, recall increasing by 0.035, F-measure increasing by 0.043, and AUC increasing by 0.056. The introduction of the attention mechanism and LSTM illustrated the correlation between code smell detection and natural language processing. Compared to the model of feature envy detection in Liu et al. [51], this study optimized it from three aspects: modifying and expanding the model structure, introducing the self-attention mechanism, and applying a snapshot ensemble.

*Traditional Machine Learning and Deep Learning*. Menshawy et al. [60] proposed a mechanism to detect feature envy code smell using machine and deep learning techniques. The study applied 6 deep learning techniques (CNN, Long Short Time Memory (LSTM), Bidirectional LSTM (BILSTM), Gated Recurrent Unit (GRU), Bidirectional Gated Recurrent Unit (BIGRU), and Autoencoder) and 11 machine learning techniques based on code structural features. The deep learning models were implemented using TensorFlow and Keras frameworks [41]. The *CNN model* was inspired by an image classification model. The CNN comprised of an input layer that passed the input features to the embedding layer. The role of this layer was to map vocabularies in high-dimension space to vectors of fixed size. The embedding output was fed to the convolution one-dimensional layer of a specific kernel and filter parameters. The new weights were computed to the next max pooling one-dimensional layer which reduced the dimensionality of the feature space. To avoid overfitting, the weights were passed to a dropout layer to randomly disregard a specific percentage of nodes during training. The weights were connected to a flattened layer and then a stack of three dense layers to predict if a given instance was smelly or belonged to the non-smelly data. The *LSTM and GRU architectures* were inspired by a typical NLP model. The input layer fed the next embedding layer with input textual features. The embedding layer mapped the input tokens to vectors. The new vectors were passed to the LSTM layer in the LSTM model or the GRU layer in the GRU model. Both networks (LSTM and GRU) had recurrent dropout and dropout values of 0.1. The new weights were fed to the dropout layer to avoid overfitting and then fed to a flattened layer and a stack of three dense layers to avoid underfitting. Similar to the CNN architecture, the input to the dense layers was one unit. The Sigmoid function was applied to decide whether the output belongs to the positive class or the negative class. A bidirectional network was applied to the LSTM and GRU layers with the same layers and hyperparameters of the LSTM and GRU models respectively. For the machine learning approach, eleven individual algorithms of different classifier families were applied. These included Decision Table, Instance-based learning with parameter K, J48, JRip, Multilayer Perception, Naive Bayes, Random Forest, Simple Logistics, Sequential minimal optimization, AdaBoost, and Bagging. Both approaches (deep learning and machine learning) were applied to open-source Java projects from the Qualitas Corpus dataset. In the machine learning approach, the Designate Java tool was used to detect the code smell and to extract the corresponding metrics in CSV files. The machine learning data processor module splits the extracted data into positive and negative samples. The machine learning algorithms were then applied to the processed output CSV samples to train the models and to evaluate the classifier's performance. In the deep learning approach, the data was tokenized by the JavaTokenizer tool which exported tokenized text files. The deep learning data processing stage splits the tokenized files into positive and negative samples according to the extracted detection information from DesignateJava. The tokenized input was then fed to the deep learning models to detect the code smells. The results showed that deep learning techniques are promising and that they tend to achieve good results compared with the machine learning approach. Based on their evaluation of the 6 deep learning techniques against machine learning models, the autoencoder models achieved superiority among all the deep learning techniques. In contrast, CNN achieved the lowest F-score. Overall, the deep learning techniques showed high potential in predicting feature envy. However, the deep learning techniques based on semantic features are not capable of detecting all code smell types.

Hamdy and Tazy [34] proposed an approach for detecting the occurrence of the god class smell in source code. Their approach utilized both the source code textual features and metrics to train three deep learning models (Long Short Term Memory, Gated Recurrent Unit, and Convolutional Neural Network). They built a dataset for the god class smell in source code acquired from the Qualitas Corpus repository. They extracted the textual features of the source code using natural language processing techniques and integrated them with metric features. They then trained the three deep learning models using different types of source code features, such as metrics, textual features, and hybrid metrics-textual features. The Convolutional Neural Network (CNN) used in the deep learning approach comprised a stack of convolution stages, for feature selection, followed by a stack of dense layers for classification. The CNN applied a set of filters on the input sequence and produced the feature map, which represented an input to the max pooling layer. The output of the last max pooling layer was connected to a dropout layer, which performed regularization by ignoring some random nodes during training to prevent overfitting. In the experiment, they set the dropout rate to be equal to 0.5. The output of the last dropout layer was fed into a dense layer, which had a fully connected multi-perceptron neural network that worked like a classifier. They used a stack of two dense layers: the first dense layer had 32 units and ReLU activation, followed by a second dense layer with the number of outputs set to equal to one. The second dense layer made predictions on whether the god class smell occurs or not in a given source code. This layer used the Sigmoid activation function to produce a probability within the range of 0 and 1. The LSTM and GRU models comprised a stack of LSTM/GRU layers, a dropout layer, and a dense layer. The LSTM/GRU layer learned the representation of each class. The regular dropout in the dropout layer was set to 0.5, while the recurrent dropout parameters of the LSTM/GRU were set to 0.1. For the machine learning approach, three traditional machine learning techniques were used: Naïve Bayes, Random Forest, and Decision tree (C4.5). These traditional approaches were mainly used to compare the effectiveness of the proposed deep learning technique. The evaluation results showed that the deep learning technique performed better than the latter.

Dewangan et al. [19] proposed a machine learning-based approach to predict code smells in software and identify the metrics that play a significant role in detecting them. They used four code smell datasets, i.e., god class, data class, feature envy, and long method, generated from 74 open-source systems (Qualitas Corpus) obtained from Fontana et al. [24]. Six different algorithms, including Naïve Bayes, KNN, Decision tree, logistic regression, Random Forest, and Multilayer Perceptron, were used in the statistical-heuristic machine learning models. The performance of each technique was evaluated individually for the four code smells. The Multilayer perceptron (MLP) algorithm was found to perform the best in terms of accuracy for detecting the data class code smell.

Barbez et al. [10] proposed a machine learning-based ensemble method called Smart Aggregation of Anti-patterns Detectors (SMAD) to create an improved classifier compared to standalone approaches. To train and evaluate the model, they created an oracle that contained the occurrences of god class and feature envy in eight open-source systems. However, neural networks often perform poorly on imbalanced datasets like their oracle, so they designed a training procedure that maximizes the expected Matthews Correlation Coefficient (MCC). They then evaluated SMAD on the oracle and compared its performance with other aggregated tools and competing ensemble methods. Although SMAD is intended for detecting antipatterns, it can serve as a benchmark for researchers looking to develop standalone tools for identifying common code smells during the refactoring process.

*4.1.4 Explainable and Feedback Centric.* This category refers to studies that have used explanation mechanisms or integrated user feedback in their approach to detecting code smells using deep learning. Explanation mechanisms typically aim to incorporate methods and techniques that facilitate the understanding and interpretability of the deep learning model's decision-making process during code smell detection. This is particularly important for developers, as it enables them to have faith in the process, especially when the detected code smells lead to critical code changes. On the other hand, user feedback integration involves actively seeking and incorporating feedback from developers into the deep learning-based code smell detection process. Continuous interaction with the developers ensures adaptability to evolving coding practices, preferences, and domain-specific knowledge.

*For the use of explainable mechanisms*, Yin et al. [104] have proposed an explainable approach to detecting feature envy based on local and global features. To make the most of the code information, they designed different representation models for global and local code. They extracted different feature envy features and automatically combined those that were beneficial for detection accuracy. They further designed a Code Semantic Dependency (CSD) to make the detection result easy to explain. The global feature contained a global semantic feature and a global metrics feature. They splice the extracted method name, enclosing class name, and target class name together to create the global semantic features. The global semantic features were metric values calculated with plugins. LSTM was used for the global semantic features to extract context from the input statements. Then, it extracted the semantic relations from context features. LSTM focused on the cell state in the neural network and used three gates to determine how much cell state information was retained. The gate structure was used to select the appropriate call state information so that LSTM could easily capture the context information between the legal name, the class name, and the target class name. For the global metrics, CNN was used to obtain complex mapping relations from simple metric information. They opted for CNN due to its ability to automatically extract features from the original features, which could reveal the relationship between metrics information and code smell. CNN is also highly suitable for parallel training on GPU which could greatly reduce the training time. To achieve the CSD, a siamese network was formed based on a local feature representation model. It had the same branch with different inputs. The code input would first pass through an embedding layer and then through an attention layer, meanwhile, the parameters were also being shared. At the final stage, they would calculate the code dependency of the feature obtained from each branch. To evaluate the approach, they used manually constructed code smell projects (Junit, PMD, JExcelAPI, Areca, Freeplane, jEdit, and Weka) and 3 real-world projects (Xmd, JSmooth, and Neuroph). They compared their approach against Liu et al.'s [53] approach, JDeodorant [22], and JMove [89]. The F-measure for the two experimental setups were 2.85% and 6.46%, respectively, which was higher compared to the state-of-the-art approaches.

*For the integration of feedback*, Nanadani et al. [64] conducted a study to investigate the effects of human feedback on the performance of trained models in detecting code smells, which are subjective perceptions of developers. To create a robust and adaptable system, the study combined deep learning techniques, user feedback, and a containerized deployment architecture for a locally-run web server. The deep learning techniques used in the study were an autoencoder with a dense multilayer perceptron, an autoencoder with a Long Short Term Memory, and a variational autoencoder with a threshold strategy for classification. The first step in this approach was to train the autoencoder, which was used to compress the input data into a lower dimensional representation called the latent representation. The autoencoder had an encoder and a decoder, with the encoder starting with an input layer followed by a series of dense layers. To improve the training stability and efficiency, batch normalization was added to standardize inputs for each mini-batch. The decoder was then constructed in the reverse order of the layers. The variational autoencoder was used to serve as a deep generative model that employed Bayesian inference to estimate the latent representation. The approach was used to detect complex method, long parameter lists, and multifaceted abstraction code smells. A plugin for IntelliJ IDEA was created, and a container-based web server was developed to offer services of the baseline deep learning model. The setup allowed developers to see code smells within the IDEA and provide feedback. Using this setup, the researchers conducted a controlled experiment with 14 participants divided into experimental and control groups. In the first round of the experiment, the code smells predicted by using the baseline deep learning model were shown, and feedback was collected from the participants. In the second round, the researchers fine-tuned and reevaluated the model's performance before and after adjustment. The results showed that calibration improves the performance of the smell detection model by an average of 15.49% in F1 score across the participants of the experimental group.

### 4.2 Code Smell Types of Refactoring Opportunities

Code smells are defined as certain structures in the code that call for the process of refactoring. Based on this definition Fowler [26] proposed 22 types of code smells, for example, long method, duplicate class, feature envy, duplicate class, etc. Code smells have been analyzed and categorized based on the implementation, design, and architectural levels (based on their scope, granularity, and impact) [26–28, 81]. Essentially, code smells may occur at different levels of the codebase, that is class, method, or variable levels. This occurrence at different levels affects how these smells may be detected when employing deep learning models. From our literature collection, we note and observe, as presented in Table 1, that a total of 26 different types of code smells have been detected using deep learning techniques which could potentially lead to the identification of refactoring opportunities. The data collected shows that the smells detected were occurring at different levels of the code base from class up to variable level. Also, we observe that some researchers focused on the detection of only one type of code smell [21, 53, 105] while others [18, 32, 48] focused on the detection of at least more than one type of code smell. Overall, our data suggest that feature envy being one of the most common code smells is the one whose detection researchers are employing the use of deep learning techniques more often. Feature envy code smell appeared in at least 27.87% of the primary studies, seconded by long method 9.84%, god class 6.56%, complex method 6.56%, and the rest of the smells. These smells, i.e., feature envy and long method could potentially lead to the recommendation and suggestion of refactorings which involve 1) moving features between objects, e.g., move method refactoring. 2) composing methods to ensure that they are much easier to understand, e.g., extract method refactoring. Through our literature search we have noted that other researchers like Xu and Zhang [102] combined several code smells to create multi smells as a way of enhancing the efficiency and validation of their approach.

### 4.3 Training Strategies

Detecting code smells using deep learning models requires the adoption of effective training strategies. These strategies involve an organized approach to teaching the model to recognize patterns, make predictions, and perform tasks based on the data. Training strategies are crucial in developing a robust and accurate model for detecting code smells. Researchers use various techniques to create effective training strategies that produce efficient models. Our analysis of the literature focuses on how researchers preprocess and engineer features from various metrics to effectively represent code smells. We also explore the embeddings and representations used to capture semantic relationships within the code, which aids the model's ability to learn code smells. Researchers use various data preprocessing, feature extraction, and data balancing techniques to enable the deep learning model to identify different code smells effectively.

*4.3.1 Data Preprocessing.* Data preprocessing is an essential step to improve the quality of data that will be fed into deep learning models. During the process of data preprocessing, several techniques are utilized, including data cleaning, data integration, data transformation, and data regularization. Based on the nature of the datasets used, various techniques may be applied to preprocess the candidate metrics such as code, text, graph data, etc. For instance, Sharma et al. [80, 81] performed their preprocessing by analyzing the data using Designate and DesignateJava on their C# and Java code, respectively. Then they split their code using Codesplit before applying tokenization. They performed duplicate removal after tokenization to ensure that no duplicate data was fed into the deep learning model. In contrast, Himesh et al. [64] conducted their duplicate removal before tokenization using a hash function to compute a unique hash value for each code instance and compared the hash values to identify any duplicates. Regularization is another widely adopted technique used during data preprocessing to prevent deep models from learning specific and irrelevant features of their data. Barbez et al. [10] used regularization to add a special term to the loss function, which encourages the weights to be small this was as defined by Witten [99]. Overall, data preprocessing is crucial in enhancing the quality and integrity of data, which in turn leads to better deep learning model performance [103].

*4.3.2 Feature Extraction.* Various features from data candidates are extracted for deep learning models to detect various code smells. The selection of features to be extracted plays a key role in the way a deep learning model makes predictions. The features that could be extracted for code smell detection include but are not limited to the following structural features, semantic features, naming conventions and documentation, patterns, etc. Several of our primary studies [18, 108, 109] were found to detect similar code smells such as feature envy, long method, god class, brain class/method, etc. respectively. However, we note that different features were employed for the same code smell across the literature. Liu et al. [51,53] leveraged the use of semantic and structural features to detect feature envy code smell but ignored the semantic information contained in input sequences. Thus, Zhang et al. [107] proceeded to include the semantic information contained in the input sequences as a way of improving the efficiency of the deep learning model. Apart from just using structural and semantic features, we note that in the detection of feature envy code smells, researchers used global metric features [104], calling relationships [105], and others even extracted ASTs from the code fragments and formed sequences of statement trees like the case of Xu and Zhang [102]. For the detection of the long method, we note that in as much as most of the researchers employed the use of source code metrics, there was a slight difference in how these were processed for them to be extracted as features. For example, Lin et al. [48] converted the metrics into XML while Hadj-Kacem and Bouassida [33] parsed the metrics into an AST. To detect god class code smell, Hamdy et al. [34] used textual features from source code while Hadj-Kacem and Bouassida [32] and Dewangan et al. [19] extracted their features from code metrics. Similarly, for brain class/method Zhang et al. [109] used both semantic and structural features while Zhang and Dong [108]and Das et al. [18] used code metrics. Table 2 gives a representative tabulation of the scenarios being highlighted. From Table 3 we note that it is only in the detection of data class where researchers used the same feature set.

Table 1. List of code smells detected in the primary studies.

| Code Smells | Frequencies | References |
|---|---|---|
| Feature envy | 17 | [10, 19, 21, 31, 33, 51, 53, 55, 60, 80, 81, 98, 102, 104, 105, 107, 111] |
| Long method | 6 | [19, 33, 35, 47, 48, 51] |
| God class | 4 | [10, 19, 32, 34] |
| Misplaced class | 1 | [51] |
| Brain class and Brain method | 3 | [18, 108, 109] |
| Complex method | 4 | [47, 64, 80, 81] |
| Complex conditional | 2 | [47, 81] |
| Multifaceted abstraction | 3 | [64, 80, 81] |
| Lazy class | 1 | [48] |
| Speculative generality | 1 | [48] |
| Refused bequest | 1 | [48] |
| Duplicate code | 1 | [48] |
| Shotgun surgery | 1 | [48] |
| Contrived complexity | 1 | [48] |
| Uncontrolled side effects | 1 | [48] |
| Insufficient modularization | 1 | [102] |
| Deficient encapsulation | 1 | [102] |
| Empty catch block/clause | 3 | [47, 80, 102] |
| Magic number | 2 | [47, 80] |
| Long identifier | 1 | [47] |
| Long statement | 1 | [47] |
| Missing default | 1 | [47] |
| Long parameter list | 2 | [47, 64] |
| Blob | 1 | [33] |
| Inconsistent method names | 1 | [54] |
| Data class | 2 | [19, 32] |
| ***Multi Smells | 1 | [47] |

Table 2. Representative feature sets.

| Code Smells | Features | References |
|---|---|---|
| Feature envy | code metrics,textual features<br>global metrics,asts | [10, 19, 21, 31, 33, 51, 53, 55, 60, 80, 81, 98, 102, 104, 105, 107, 111] |
| Long method | code metrics, source code (xml) | [19, 33, 35, 47, 48, 51] |
| God class | code metrics, textual features | [10, 19, 32, 34] |
| Brain class/method | code metrics, textual features | [18, 108, 109] |
| Complex method | source code, code metrics | [47, 64, 80, 81] |
| Complex conditional | source code, code metrics | [47, 81] |
| Multifaceted abstraction | source code code metrics | [64, 80, 81] |
| Empty catch block | source code, code metrics | [47, 80, 102] |
| Magic number | source code, code metrics | [47, 80] |
| Long parameter list | code metrics,source code | [47, 64] |
| Data class | code metrics | [19, 32] |

4.3.3 *Feature Embedding.* Deep learning models require the features extracted to be converted into a suitable format for the model to extract the relevant features. Tokenization and vectorization are common processes used to achieve this. Tokenization breaks down text into smaller units (tokens), while vectorization converts these tokens into numerical vectors. Researchers also use embedding techniques to convert the candidate data into the required format. Embedding is a specific type of vectorization that represents words as dense vectors in a continuous space, capturing semantic relationships between words. Table 3 lists some of the tools that researchers used to execute these processes. Our data indicates that most researchers used Word2Vec as an aid to conducting embedding for their data candidates, with 23.08% of the primary studies using it, followed by iplasma at 19.23%. We also discovered that at least 15% of the studies did not explicitly highlight the tools used for the feature extraction processes.

4.3.4 *Data Balancing.* Data balancing is the process of ensuring that the different categories or labels in a dataset are represented equally in terms of their frequency. This is particularly important in classification problems where the model tries to predict different classes. A balanced dataset prevents the model from being biased towards the majority class and ensures that it performs well in all classes. Imbalanced datasets, on the other hand, can lead to poor performance of the model, especially for the minority class. The detection of code smells also requires a balanced dataset where the number of positive samples (with code smell) and negative samples (without code smell) are maintained. Different techniques can be used to balance the dataset such as under-sampling or oversampling. Under-sampling removes samples from the majority classes while oversampling generates new samples for the minority class. Synthetic data generation and ensemble methods can also be used to reduce the disadvantages of having an imbalanced dataset. Researchers have used the Synthetic Minority Over Sampling Algorithm (SMOTE) to balance their data in the detection of code smells. SMOTE is an effective oversampling technique that has been widely used in practice. However, its effectiveness may vary depending on the specific characteristics of the data and the problem at hand. It is worth noting that while 78.5% of the primary studies were clear about the techniques they used to balance their data, 21.5% did not mention any issues with data imbalance.

Table 3. Feature extraction tools.

| Tools | References |
|---|---|
| Word2Vec | [51, 53, 54, 102, 104, 109] |
| iplasma | [18, 31, 32, 108, 109] |
| Javalang | [34] |
| NLTK | [34] |
| Tokenizer | [80, 81] |
| Fluid tool | [31, 32] |
| Javatokenizer | [60] |
| GraphSAGE | [105] |
| Wrapper | [19] |
| CodeT5 | [55] |

*4.3.5 Model Training.* Various approaches exist for training deep learning models. According to our literature collection, models can undergo training using labeled data, unlabeled data, or a blend of both. The process of labeling data can occur automatically, semi-automatically, or manually. These training methodologies are commonly categorized as supervised, unsupervised, or a combination of both, where a simultaneous application of both techniques is employed. These techniques are presented in Figure 3. Our survey found that the majority of primary studies (65.38%) used supervised learning techniques, while 23.08% used unsupervised techniques and 11.54% used a combination of both.

### 4.4 Evaluation

*4.4.1 Datasets.* Based on our survey findings, a majority of the primary studies utilized applications exclusively developed in a single programming language, such as Java, for conducting experiments and evaluations. In contrast, some studies opted for a diverse approach by employing a combination of applications developed in different languages, such as Java and C#, for their experiments and performance assessments. As depicted in Figure 4, the data reveals that 87.5% of the primary studies relied on projects developed in a singular programming language, while only 12.5% employed a combination of programming languages in their application projects. Additionally, as depicted in Figure 5, a significant majority, specifically 91.67%, of the studies employed projects that are publicly accessible, i.e., general open-source projects available in various repositories, and real-world open-source projects. In contrast, 8.33% utilized private projects with restricted access.

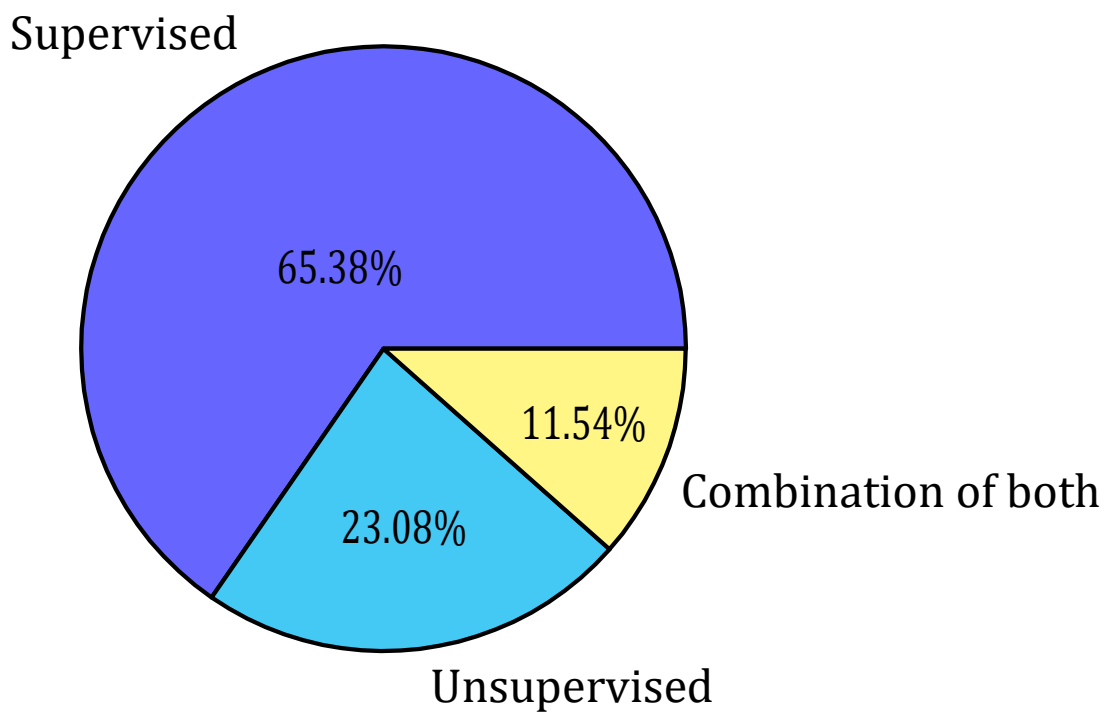

Fig. 3. Model training strategies

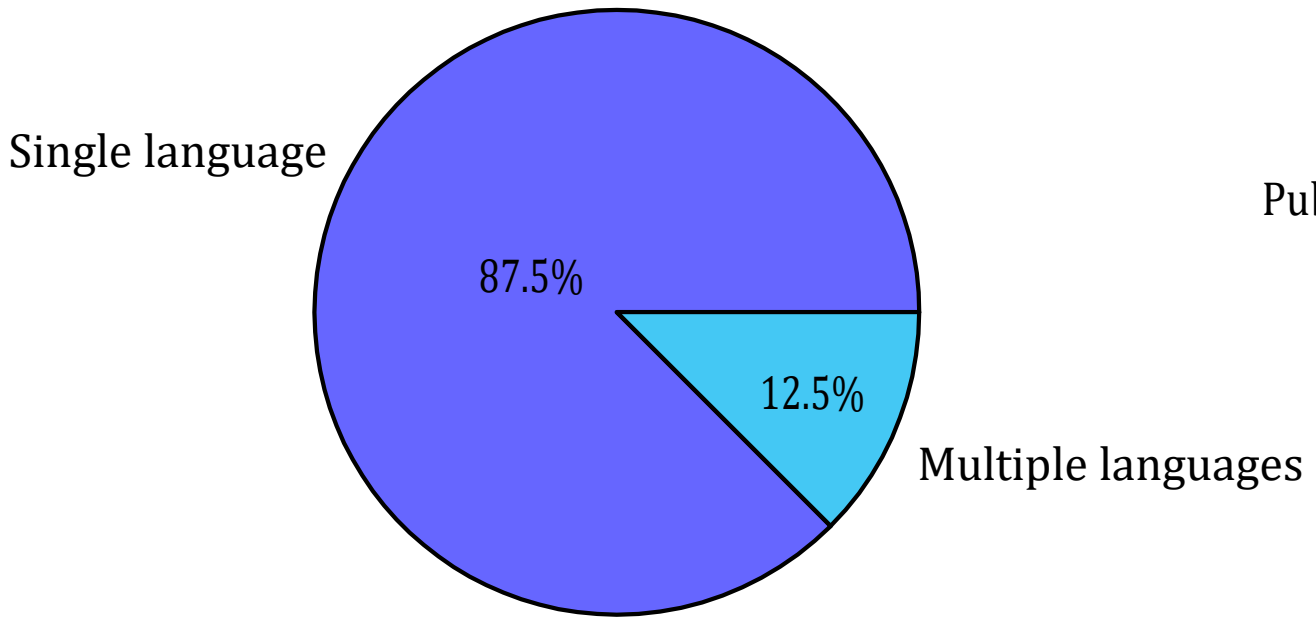

Fig. 4. Programming languages

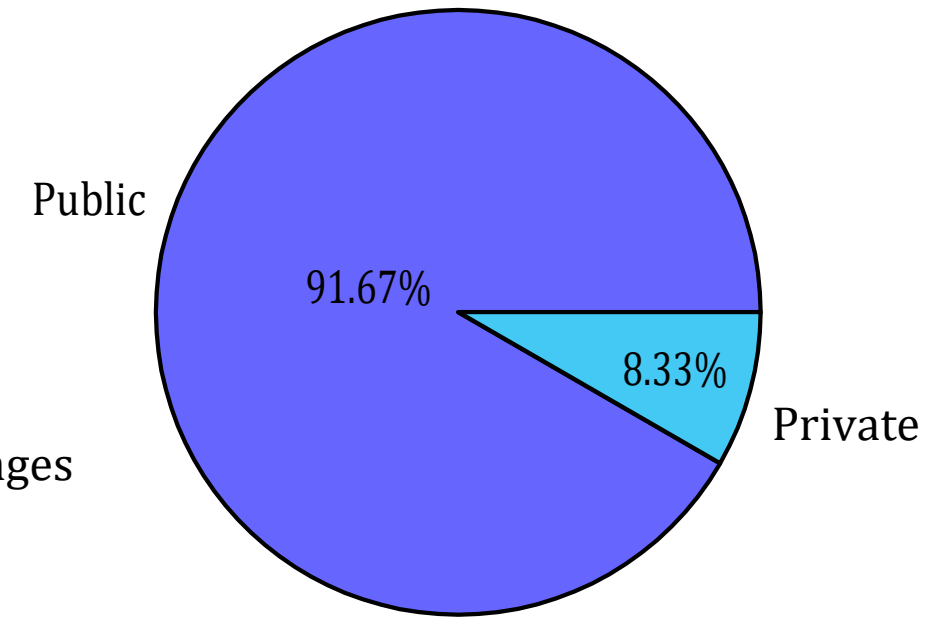

Fig. 5. Dataset types

*4.4.2 Result Metrics.* Researchers use metrics to evaluate the performance of a particular approach. Commonly used metrics for this calculation include precision, recall, F-measure, and accuracy. Precision measures the accuracy of the positive predictions. Its formula is defined as (Precision = TP / (TP + FP)), where TP represents true positive and FP represents false positive. It is important in situations where false positives should be minimized, and there is a cost associated with making incorrect positive predictions. Recall measures the model's ability to capture all relevant cases. Its formula is defined as (Recall = TP / (TP + FN)), where FN represents a false negative. It is important when the goal is to capture as many positive cases as possible, and there is a cost associated with missing positive predictions. The F1 score or F-measure provides a balance between precision and recall. Its formula is defined as (F1 Score = 2 * (Precision * Recall) / (Precision + Recall)). It is useful when both false positives and false negatives need to be minimized. Accuracy measures the ratio of correctly predicted observations to the total observations. Its formula is defined as (Accuracy = (TP + TN) / (TP + TN + FP + FN)), where TN represents true negative. While accuracy is a common metric, it may not be suitable for imbalanced datasets as it can be misleading. Accuracy is widely used when the classes are balanced. However, in imbalanced datasets, accuracy alone may not provide a complete picture of model performance. From our primary studies, we have noted that researchers use other metrics apart from the outlined (precision, recall, f-measure, and accuracy). For example, researchers [35, 51, 80, 107] have used Area Under Cover(AUC) to represent the trade-off between true positive rate and false positive rate at various thresholds. For instance [10, 51], used the Matthews Correlation Coefficient (MCC). MCC is a correlation coefficient between the observed and predicted binary classifications. It takes into account true and false positives and negatives and is particularly useful when dealing with imbalanced datasets. Its formula is defined as (MCC = (TP * TN - FP * FN) / sqrt((TP + FP) * (TP + FN) * (TN + FP) * (TN + FN))). MCC is a balanced metric that works well for imbalanced datasets. Lin et al. used [48] Kappa (Cohen's Kappa). Kappa measures the agreement between observed and predicted classifications, correcting for the agreement that could occur by chance. Its formula is defined by ( Kappa = (Po - Pe) / (1 - Pe)), where Po is the observed agreement, and Pe is the expected agreement. Kappa is useful for classification problems where class distribution is imbalanced, and it accounts for the agreement that could occur by chance. From Figure 6 it becomes apparent that precision, recall, and F-measure are commonly used for performance assessment. Notably, 96.15% of our primary studies utilized precision, while 92.31% and 88.46% employed recall and F-measure, respectively.

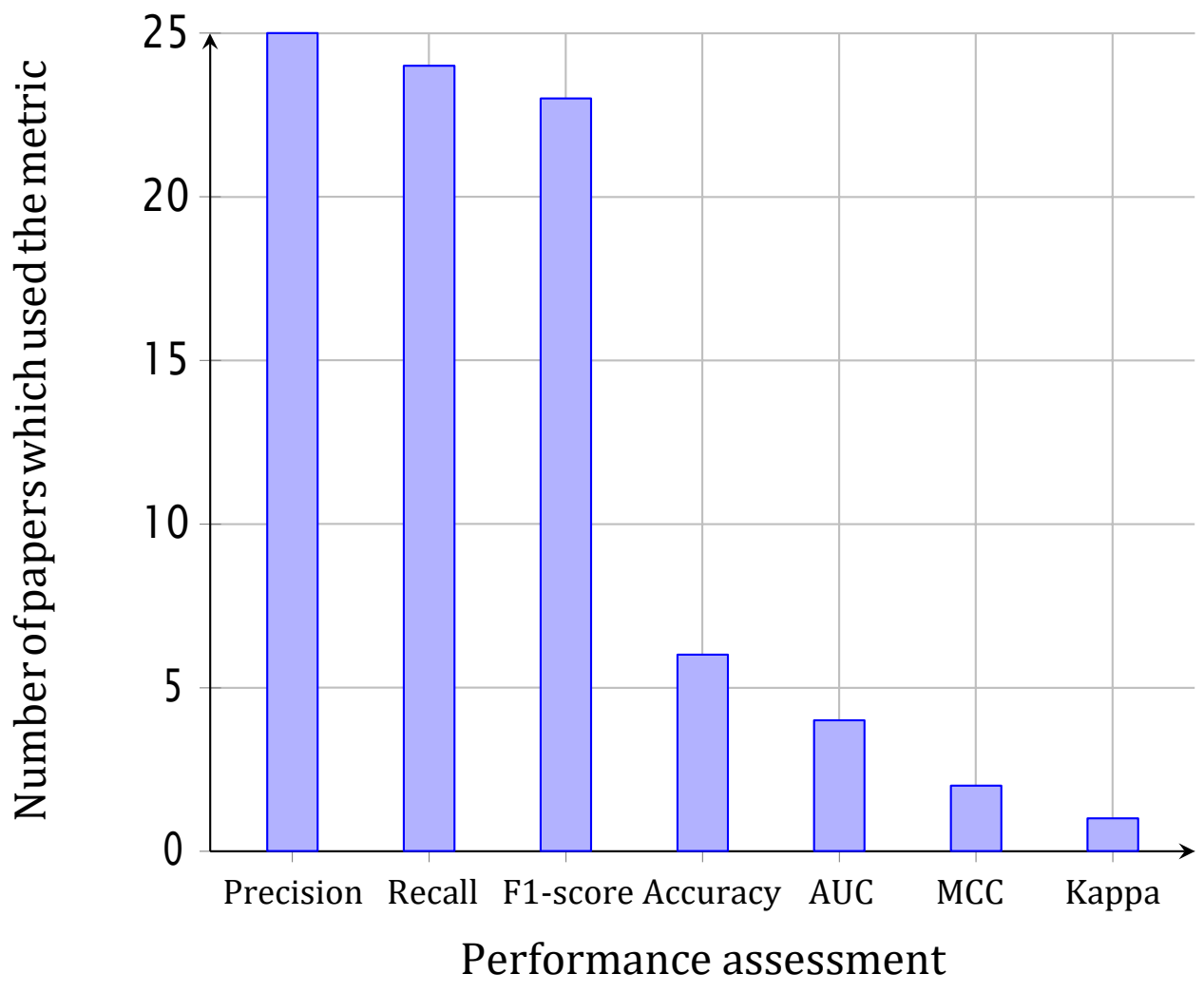

Fig. 6. Metrics used for performance assessment

## 5 RECOMMENDATION OF REFACTORING SOLUTIONS

The recommendation of refactoring solutions involves identifying areas of code that could benefit from refactoring and suggesting potential approaches or strategies for improving them. Recommendations are often made based on code reviews, automated analysis tools, or architectural discussions. The recommendation of refactoring solutions does not involve directly altering the code but rather advising on possible improvements. The recommendation of refactoring solutions is more about identifying problems and proposing solutions than actually implementing them. In this section, we will explore how deep learning techniques have been used by researchers to suggest refactoring solutions. We will discuss the recommendation technologies, the suggested refactorings, the training strategies, and the evaluation metrics used to assess the performance of the approach.

### 5.1 Recommendation technologies

Through the use of various deep-learning techniques, researchers have been able to suggest solutions for refactoring. From our literature search, we note that some researchers [3, 9, 44] have utilized deep learning models to predict specific refactoring solutions. Conversely, others [51, 53, 55] have utilized deep learning models to initially detect the presence of code smells in a given codebase and subsequently recommend refactoring solutions based on the identified code smells. Consequently, we have categorized these approaches into two distinct groups: refactoring-based and code smell-based. We will discuss the technologies used in these studies in the following sections.

*5.1.1 Refactoring Based Technologies.* Alenezi et al. [3] conducted a study on the effectiveness of deep learning algorithms in building refactoring prediction models at the class level. They used a Gated Recurrent Unit (GRU) as a key structural parameter. The GRU comprised a hidden layer, an epoch, and a normalization batch. The GRU had three hidden layers and utilized the number of epochs between 10 to 2500, with a batch size of 2 to 25. The deep learning algorithm proposed by Qasem et al. [76] was used after the three hidden layers. The study used 7 open-source Java-based projects to assess the effectiveness of the proposed algorithm, which had two main stages. In the first stage, a set of necessary preprocessing procedures were performed on the datasets. During this initial stage, SMOTE was used to prevent any imbalance. In the second stage, the deep learning algorithm was applied to the dataset to predict the need for refactoring at the class level by using the Gated Recurrent Unit algorithm. For the experiment, two sets of datasets, balanced and unbalanced, were used to check if the balance of the dataset would affect the predictability of the deep learning model. The results indicated that a balanced dataset enhances the prediction of class-level refactorings.

Aniche et al. [9] conducted a large-scale study to evaluate the effectiveness of various supervised machine learning algorithms in predicting software refactoring. They aimed to demonstrate that machine learning methods can accurately model the refactoring recommendation problem. The study used a dataset containing over two million real-world refactorings extracted from more than 11,000 real-world projects from Apache, F-Droid, and GitHub repositories. The study evaluated six machine learning algorithms, of which five were traditional machine learning approaches (logistic regression, naive Bayes, support vector machine, decision trees, and random forest), and one was a neural network used as the deep learning technique. The neural network used was a sequential network of three dense layers with 128, 64, and 1 units, respectively. To avoid overfitting, dropout layers were added between the sequential dense layers, keeping learning in 80% of the units in dense layers. The number of epochs was set to 1000. All the algorithms employed for this technique were then individually used to predict refactorings at different levels: class, method, and variable. The resulting models were able to predict 20 refactorings at the different levels with an accuracy higher than 90%.

Kumar et al. [44] developed a recommendation system that suggests which methods require refactoring. They used 25 different source code metrics at the method level as input features in a machine learning framework. The system then predicts the need for refactoring. The authors conducted a series of experiments on a publicly available annotated dataset of five software systems to investigate the performance of the approach. The approach used ten different machine learning classifiers, of which seven (AdaBoost, LegitBoost, Log, NB, BayesNet, RF, and RBFN) were traditional machine learning techniques. Only three out of the ten were based on deep learning, namely MLP, ANN-GD, and ANN-LM. The first phase of the approach focused on analyzing relevant features. The considered dataset was preprocessed to extract relevant features. At this stage, the Wilcoxon rank sum test was applied to handle uncertainty in the dataset, and also for extracting features. Subsequently, ULR analysis was applied to identify the final set of source code metrics considered for further implementation. The second phase involved the model-building process and the assessment of the performance of the proposed model. During the model-building process, the dataset was normalized through min-max normalization. The class imbalance issue of the dataset was addressed using SMOTE, RUSBoost, and Up-Sample. Then, a 10-fold cross-validation was employed with the 10 machine learning techniques to implement the proposed approach. The results obtained from each technique were compared with different performance measures to evaluate them. The authors concluded that overall, from the experiments conducted, method-level refactoring prediction using source code metrics and machine learning classifiers is possible.

Panigrahi et al. [69] designed a recommendation system for predicting refactoring instances at the class level using machine learning techniques, various data sampling approaches to address the data imbalance problem, and feature selection techniques. The approach centered around five main research questions: 1) how does data balancing improve the prediction model's capability? 2) how does feature selection improve the results of the refactoring prediction model? 3) how do machine learning classifiers predict the refactoring model's results across different performance parameters? 4) how can refactoring instances be predicted in the case of cross-project? 5) how can refactoring instances be predicted in the case of intra-project? The authors highlighted the importance of using machine learning to refactor prediction models for large object-oriented software systems. They recommended using ensemble approaches and enhanced machine learning classification algorithms to predict superior performance across different performance parameters. The classifiers recommended for use as standalone or ensemble classifiers include SVM, LSSVM, Naïve Bayes, Random Forest, ELM-based, K-nearest-neighbors, and deep learning models such as LSTM and CNN. For the approach in this study, the authors used LSTM and CNN classifiers. The first phase involved calculating software metrics using the source meter tool, while the second phase involved data normalization. Relevant features were selected from all the features extracted from the dataset using Principal Component Analysis (PCA), correlation tests, and Wilcoxon rank tests. The classifiers were then utilized to predict refactoring instances. To improve the refactoring model's performance, they implemented ensemble techniques, LSTM, and CNN. Performance was evaluated using different performance metrics such as recall, accuracy, F-measure, and precision. For unbalanced data, the Area Under the Curve (AUC) was computed.

Sagar et al. [79] conducted research to determine whether code metrics can be used to predict refactoring activities in the source code. They approached this by formulating refactoring operation type prediction as a multi-classification problem and implementing both supervised learning and LSTM models. They used the metrics extracted from committed code changes to extract the features that best represent each class and predict the method level refactoring being applied ( move method, rename method, extract method, inline method, pull up method, and push down method) for any project. For this approach, they developed two types of models - a metric-based model that included traditional machine learning techniques (random forest, SVM, and logistic regression classifiers) and a text-based model with LSTM. The *text-based model* had an input layer of word embedding metrics and an LSTM layer. The LSTM layer provided the final dense layer output. For the LSTM layer, they used 128 neurons for the dense layer, and five neurons since there were five different refactoring classes. They used Softmax as an activation function in the dense layer and categorical cross-entropy as the loss function. The *metric-based model* was built with supervised machine learning models to predict the refactoring class. The random forest, SVM, and logistic regression were trained with 70% of the data. Initially, the proposed metric model was implemented with only commit messages as input, but the authors realized that this approach was insufficient. Therefore, they combined commit messages with code metrics in the second experiment. The model built with LSTM produced 54.3% accuracy. The model built with sixty-four different code metrics dealing with cohesion and coupling characteristics of the code produced 75% accuracy when tested with 30% of data. The study showed that commit messages with little vocabulary are not sufficient for training machine learning models.

Cui et al. [17] have proposed an approach for recommending move method refactoring called Rmove. The proposed approach involves automatically learning both structural and semantic representations from code snippets. To achieve this, they first extracted method structural and semantic information from a dataset. Next, they created the structural and semantic representation and concatenated them. Finally, they trained a machine learning classifier to guide the movement of the method to a suitable class. The authors used a total of nine classifiers, of which six (Decision Tree, Naïve Bayes, SVM, Logistic Regression, Random Forest, and Extreme Gradient Boosting) were machine learning-based. Only three were based on deep learning, namely CNN, LSTM, and GRU. The approach demonstrated significant improvement, with an increase of 14%-36% in precision, 19%-45% in recall, and 27%-44% in F-measure compared to the state-of-the-art techniques, such as PathMove [45], JDeodorant [22], and JMove [89].

Nyamawe et al. [66] proposed a learning-based approach for recommending refactoring types based on the history of feature requests, code smells information, and the applied refactorings on the respective commits. The proposed approach learned from the training dataset associated with a set of applications. The approach could be used to suggest refactoring types for feature requests associated with other applications or that are associated with the training applications. The proposed approach had six main steps. Firstly, the feature requests were extracted from the issue tracker JICA and their respective commits were retrieved from a repository on GitHub. Secondly, the previously applied refactorings on the retrieved commits were recovered. Thirdly, RefDiff [84] and RMiner [95] were used to identify the code smells associated with the source code in each of the retrieved commits. Fourthly, text processing was applied to the contents of the file to prepare textual data into a numerical representation for training the classifiers. The fifth step involved the training of the classifiers which gave the prediction model for predicting and recommending refactorings for new feature requests. Six classifiers were employed, out of which five (SVM, MNB, LR, RF, and DT) were machine learning-based, while only one was deep learning-based (CNN). The study noted that CNN performed slightly lower than the rest of the classifiers, partly because deep learning classifiers generally require significantly larger datasets to achieve competitive performance. Nonetheless, the overall evaluation of the approach based on two tasks (the need for refactoring and recommending refactoring types) indicated that the approach attained an accuracy of up to 76.01% and 83.19%, respectively.

In a similar vein to the work done in Nyamawe et al. [66], Nyamawe [65] proposed a machine learning approach that made use of commit messages to improve software refactoring recommendations. This approach identified past refactorings that were applied to commits used for implementing feature requests by analyzing the commit messages. The approach employed six algorithms, including five based on machine learning (SVM, MNB, Random Forest, Logistic Regression, and Decision Tree) and one based on deep learning (CNN). To evaluate the approach, a dataset of commit messages from 65 open-source projects was used. The results showed that leveraging commit messages improved refactoring recommendation accuracy significantly compared to the state-of-the-art.

A study conducted by Mastropaolo et al. [59] explored the potential of data-driven approaches to automate variable renaming. They experimented with three techniques - a statistical language model and two deep learning-based models. Three datasets were used to train and evaluate the models. The first dataset was used to train the models, tune their parameters, and assess their performance. The second and third datasets were used to further evaluate the performance of the techniques. The researchers found that under certain conditions, these techniques can provide valuable recommendations and can be integrated into rename refactoring tools. The three representative techniques used were a statistical model, an n-gram cached language model proposed by Hellendoorn [36], T5 proposed by Raffel et al. [77], and a transformer-based model presented by Liu et al. [50]. The study demonstrated that deep learning models, particularly those that generate predictions with high confidence, can be valuable support for variable rename refactoring.

Alomar et al. [4] developed a tool called AntiCopyPaster, which is a plugin for IntelliJ IDEA. The tool aims to provide recommendations for extract method refactoring opportunities as soon as duplicate code is introduced in the opened file in the IDE. The tool takes into account various semantic and syntactic code metrics as input and makes a binary decision on whether the code fragment should be extracted or not. The goal of this approach is to increase the adoption of extract method refactoring while maintaining the workflow of the developer. To achieve this goal, Alomar et al. in [5] investigated the effectiveness of machine learning and deep learning algorithms. They defined the detection of extract method refactoring as a binary classification problem. Their proposed approach relied on mining prior applied extract method refactorings and extracting their features to train a deep learning classifier that detected them in the user's code. The approach was structured into four phases: data collection, refactoring detection, code metrics selection, and tool design and evaluation. The deep learning model used in the approach was CNN. The CNN comprised multiple layers of fully connected nodes, structured into convolutional, deconvolutional, and dense layers. A dropout stage was also included to prevent overfitting. The input to the CNN was a vector of 78 metric values which were batch-normalized to stabilize their distribution. The batch normalized inputs were then fed into a convolution that reduced the feature space from 78 to 32. ReLU was used as the activation function for the convolutional layers. The convoluted data was then fed into the deconvolutional layer, which was followed by a max pooling layer with a filter size of 2 that took the largest number in the filter. The final layer was the dense layer, in which each node received input from all nodes of the previous layer. The approach was implemented as a plugin in IntelliJ IDEA, which is a popular IDE for Java. The plugin consists of four main components: Duplicate detector, Code analyzer, Method extractor, and Refactoring launcher. The results showed that CNN recommended the appropriate extract method refactorings with an F-measure of 0.82. These results solidify that machine learning models can recommend extract method refactorings while maintaining the workflow of the developer.

Cui et al. [16] have proposed an automated approach called Representation-based Extract Method Refactoring Recommender System (REMS) to suggest appropriate extract method refactoring opportunities. The approach involves mining multi-view representations from a code property graph. First, code property graphs were extracted from training and testing samples. Then, multi-view representations such as tree-view and flow-view representations were generated from the code property graph. Compact bilinear pooling was used to fuse the tree-view and the flow-view representations. Finally, machine learning classifiers were trained to guide the extraction of suitable lines of code as a new method. Six relevant embedding techniques such as CodeBERT, GraphCodeBERT, CodeGPT, CodeT5, PLBART, and CoTexT were used to generate various representations of the abstract syntax tree, which were referred to as tree-view representations. The researchers explored the impact of these representations on recommendation performance. The REMS operates in three phases, including 1) feature extraction from code property graphs of training and testing samples, 2) model training based on machine learning techniques, and 3) applicable analysis of behavior preservation and functional usefulness. Seven traditional machine learning models such as Decision Tree, K-nearest neighbor, Logistic Regression, Naïve Bayes, Random Forest, Support Vector Machine, and Extreme Gradient Boosting, were used. CNN and LSTM were the only deep-learning techniques employed. The results showed that the REMS approach outperformed five state-of-the-art refactoring tools, including GEMS [101], JExtract [83], SEMI [14], JDeodorant [22], and Segmentation [90].

Pantiuchina [71] has developed techniques to create a new generation of refactoring recommenders. These recommenders can predict code components that are likely to be affected by code smells in the near future and recommend meaningful refactorings that emulate the ones that developers would perform. They refer to this approach as just-in-time rational refactoring, which has two main goals. First, predicting code quality decay aims to develop techniques that alert the developer when a code component is deviating from good design principles before design flaws are introduced. Second, learning refactoring transformations investigates the possibility of applying deep learning models to learn code changes performed by software developers. The researchers plan to investigate if neural machine translation models can be used for the replication of refactoring operations performed by software developers.

Pinheiro et al. [74] investigated how trivial class-level refactorings could affect the prediction of non-trivial refactorings using machine learning techniques. They selected 884 open-source projects and extracted the type of refactoring from classes involved in some operation and code metrics. The researchers grouped the refactorings into trivial and non-trivial ones based on their level of change. Trivial refactorings included adding class annotations, changing access modifiers, removing class annotations, etc. Non-trivial refactorings included extract class, move class, merge class, etc. Additionally, they proposed contexts based on combinations of the refactoring types that made it possible to increase the accuracy of supervised learning models. They used four traditional machine learning models, Decision Tree, Logistic Regression, Naïve Bayes, and Random Forest. They employed a Neural Network as the only deep-learning technique for this approach. They followed a sequence of five steps: selection of software projects, refactoring, and feature mining, selection of contexts, training and testing of the machine learning-based models, and evaluation of the results. The selection of contexts in step number three had to do with creating several datasets with different combinations of refactoring types. The datasets constructed by the combinations of C1, C2, and C3 were used to predict the refactorings. The four machine learning models were used through the Scikit-learn library, while the Neural Network was used through Tensorflow Keras. After training, each generated model was validated by predicting the refactorings of the features in the test set. The main findings of this approach were: 1) machine learning with tree-based models, such as Random Forest and Decision Tree, performed very well when trained with code metrics to detect refactorings, 2) separating trivial and non-trivial refactorings into different classes resulted in a more efficient model, indicative to improve the accuracy of machine learning-based automated solutions, and 3) using balancing techniques that increase or decrease samples randomly is not the best strategy to improve datasets composed of code metrics.

Panigrahi et al. [70] developed a refactoring prediction model using an ensemble-based approach. They identified the optimal set of code metrics and their association with refactoring proneness by analyzing the structural artifacts of the software program. This approach involved refactoring data preparation, feature extraction, multiphased feature extraction, sampling, and heterogeneous ensemble structure refactoring prediction. The proposed model extracted 125 software metrics from object-oriented software systems using a robust multi-phased feature selection method, which included Wilcoxon significant text, Pearson correlation test, and Principal Component Analysis (PCA). The optimal features characterizing inheritance, size, coupling, cohesion, and complexity were retained. A novel heterogeneous ensemble classifier was developed using techniques such as ANN-Gradient Descent, ANN-Levenberg Marquardt, ANN-GDX, ANN-Radial Basis Function support vector machine, LSSVM-Linear, LSSVM-Polynomial, LSSVM-RBF, Decision Tree algorithm, Logistic Regression algorithm, and Extreme Learning Machine (ELM) model as the base classifiers. The results indicated that the Maximum Voting Ensemble (MVE) achieved better accuracy, recall, precision, and F-measure values (99.76, 99.93, 98.96, 98.44) compared to the Base Trained Ensemble (BTE). Additionally, it experienced fewer errors (MAE = 0.0057, MORE = 0.0701, RMSE = 0.0068, and SEM = 0.0107) during the implementation to develop the refactoring model. The experimental results recommended that MVE with up-sampling could be implemented to improve the performance of the refactoring prediction model at the class level.

*5.1.2 Code Smell Based Technologies.* Apart from proposing a deep learning-based approach to identify feature envy smells, one of the most common code smells, Liu et al. [53], also used their approach described in Section 4 to recommend move method refactorings. For methods that were predicted to be smelly (with feature envy), they suggested that such methods should be moved via move method refactorings. If only one (noted as *inputj*) of the testing items generated for a method *m* was predicted as positive, they suggested moving *m* to the target class *tcj* that was associated with the positive testing item *inputj*. If more than one testing items were predicted as positive, they selected one (noted as *inputi*) with the greatest output and suggested moving method *m* to class *tci* that associated with *inputi*. Although their neural network described in Section 4 was a binary classifier the output of the neural network was a decimal varying from zero to one. The neural network interpreted the prediction as positive if and only if the output was greater than 0.5. [15]. Their results indicated that the approach was accurate in recommending destinations for the smelly methods. The approach achieved, on average, an accuracy of 74.94%. They also observed that the approach was more accurate than the state-of-the-art tool in the recommendation of move method refactorings, i.e., JMove [89], JDeodorant [22]. This study was extended in Liu et al. [51] where in addition to using deep learning to detect code smells they also explored the recommendation of refactoring solutions for feature envy and misplaced class. The recommendation of refactoring solutions for feature envy was the same as in Liu et al. [53]. The CNN used for misplaced class was the same as described in Section 4. To decide whether a given class should be moved from its enclosing package to another package, they leveraged two categories of features code metrics and textual features. The used code metrics included coupling between objects and message-passing coupling. To evaluate the approach they compared their proposed approach to TACO [68] in recommendation of target packages for misplaced classes. The accuracy of the recommendation was critical because if misplaced classes are moved to the wrong positions they remain misplaced. The approach resulted in a greater number of accepted recommendations. In total, 488 of its recommendations were accepted whereas the number was reduced to 342 for TACO [68]. Secondly, TACO [68] was more accurate than the proposed approach in recommending target packages. It improved the average accuracy from 49.80 to 62.98 percent. However, on the same code smells where both the proposed approach and TACO [68] made recommendations, their accuracy in recommending target packages was close to each other 62.6% for TACO [68] and 60.05% for the proposed approach. Based on this analysis they concluded that the proposed approach outperformed the baseline in identifying misplaced classes and it could be comparable to the baseline in recommending target packages.

Ma et al. [55] in their pursuit of using pre-trained model CodeT5 to extract the semantic relationship between code snippets to detect feature envy code smell, also explored the effectiveness of their approach in recommending refactoring solutions for the code smell. They wanted to find out if their approach could exceed the state-of-the-art approaches in recommending destinations for the methods to be moved. In their approach, for methods that were predicted as smelly (with feature envy, *inputec* was predicted positive), they suggested where such a method should be moved via move method refactoring. Then, they fed all testing items *inputptci*=<*code(m),code(ptci)*> into the neural network. If only one *(noted as inputptci)* of the testing items generated for m is predicted as positive, then they suggested moving m to the potential target class *(ptcj)* that is associated with the positive testing item *inputptcj*. If more than one testing item is predicted as positive, they selected the one (noted as *inputptci*) with the greatest output and suggested moving method m to class *ptci* that is associated *inputptci*. The neural network interpreted the prediction as positive if and only if the output was greater than 0.5. This approach was compared to Liu et al.'s approach [51]. The results indicated that their approach improved the accuracy in recommending target classes as it attained an accuracy rate of greater than 90% while Liu et al.' was below 90%.

The Bi-LSTM with a self-attention mechanism that was proposed by Wang et al. [98] to detect feature envy code smell was also used to recommend refactoring destinations for the methods to be moved. For a method *m* that had been marked as positive, they predicted its refactoring destination as follows. If there is only one positive example, they regard the target class related to this example as the destination. If there was more than one, they chose the target class related to the highest probability in the output set. The rationale behind this was that higher probability meant higher confidence in the deep neural network obtained. This solution was presented as *destination*=*Cmax where Cmax* related to *Pmax*=*Maxp1,p1,. ,pk*. This functional mapping of input to output was presented and computed by the deep learning model. This approach was compared to JMove [89], JDeodorant [22], and Liu et al.'s [53] approach. The results indicated that their approach was more accurate on destination recommendation than the state-of-the-art.

Yu et al. [105] did not only solve the problem of inherent calling relationships between methods which usually cause unimpressive detection efficiency by proposing a Graph Neural Network (GNN) based approach towards feature envy detection but also utilized the strength of the calling relationship of one method to another to recommend refactoring solutions for the feature envy code smell. For the methods that were predicted to have a smell, they provided refactoring suggestions to move these methods to the classes that best fit their functional implementation. If a smelly method accessed only one external class, they recommended moving it to that external class. If the smelly method was interested in two or more external classes, they employed an algorithm to suggest the most suitable external class. For this algorithm, they regarded the calling strength as the weight of the edge and obtained the calling strength graph. The calling strength graph was constructed as G2=V, E, W where V represented the set of nodes, E represented the set of edges and W represented the weight matrix of edges. To validate their approach, they compared it against, JDeodorant [22], JMove [89], and Liu et al.'s [51] approach. The results indicated that their approach had higher accuracy than the three benchmarks in refactoring recommendations. Specifically, compared with Liu et al.'s [51] work, JDeodorant [22], and JMove [89], it improved the accuracy by 10.10%, 5.13% and 11.00% respectively.

Liu et al. [54] proposed an automated approach to detecting and improving inconsistent method names. In addition to identifying inconsistent names, their approach provided a list of ranked suggestions for new names for a given method. The ranked list of similar names was generated using four ranking strategies. The first strategy (R1) relied solely on the similarities between method bodies, ranking the names of similar method bodies according to their similarity to the given method body. The second strategy (R2) grouped identical names, ranked distinct names based on the size of the associated groups, and broke ties based on the similarities between method bodies as per R1. The third strategy (R3) was similar to R2, but it ranked groups based on the average similarity, regardless of the group size. To avoid highly-ranked but small groups, the fourth strategy (R4) re-ranked all groups produced in R3, downgrading all 1-size groups to the lowest position. To evaluate the performance of their approach in suggesting new names for inconsistent names, the suggested names were ranked using the aforementioned strategies. To ensure a comprehensive assessment, three different scenarios were considered: inconsistency avoidance, first token accuracy, and full name accuracy. The approach achieved an accuracy of 34-50% in suggesting subtokens and 16-25% accuracy in suggesting full names.

Liu et al. [21] conducted a study to enhance the deep learning-based feature envy detection approaches by providing real-world examples. They used their approach to identify feature envy methods that should be moved from their enclosing classes to other classes that they envy. They achieved this by using a heuristic-based filtering method, as outlined in Section 4. Their approach, feTruth, utilized a trained classifier and a sequence of heuristic rules to predict whether a given method in the testing project was associated with feature envy smell. If the method was associated with feature envy smell, feTruth would suggest the class to which the method should be moved. The accuracy of feTruth was compared against other approaches, namely JDeodorant [22], JMove [89], and Liu et al.'s [51] approach. The results showed that feTruth was accurate in suggesting destination classes for feature envy methods with an accuracy of 93.1%. This was higher than JDeodorant's accuracy of 80% and comparable to Liu et al.'s [51] and JMove's accuracy of 87.5% and 100%, respectively.

### 5.2 Refactoring Types

From the primary studies presented in this section, we note that various refactoring solutions were recommended as a way of addressing issues related to a particular codebase. When recommending a refactoring solution, a particular refactoring that could be applied is suggested to resolve the identified issue. Refactorings are techniques that are applied to the codebase to improve its quality without altering its external behavior. Refactorings are usually applied at different levels of the codebase namely class, method, and variable. Refactorings applied at the class level are usually aimed at enhancing the overall structure, maintainability, and readability of the codebase by making changes at the class level. Examples of these include extract class, collapse hierarchy, rename class, etc. Method-level refactorings involve modifying the internal structure of methods to enhance readability, maintainability, and performance without altering the external behavior of the code. The goal of these is to create cleaner, more efficient, and easier-to-understand methods. Examples include extract method, rename method, move method, inline method, etc. Variable-level refactorings involve making changes to the variables within the codebase to improve the quality, readability, and maintainability without altering the external behavior of the program. Examples of these include rename variable, extract variable, inline variable, encapsulate field, etc.

Table 4 presents the level of the refactoring solutions that were recommended by the primary studies in our survey plus the representative refactorings suggested at that level. Analyzing the presented data, it becomes apparent that a majority of the recommended refactoring solutions focus on the method level, with 58.06% of researchers proposing solutions at this granularity. Class-level refactoring follows at 25.81%, and variable-level solutions are suggested in 16.13% of the primary studies. Through this data, we also observed that under the method level refactoring, the most common refactoring types being suggested were the extract method and the move method. We noted that extract method refactoring was found in 62.50% of the studies that employed refactoring based recommendation approach while move method was found in 85.71% of the studies that employed code smell-based approach to recommendation refactoring.

### 5.3 Training Strategies

As previously mentioned, training strategies play a crucial role in developing a robust and accurate deep-learning model. In this section, we will discuss the technicalities, tools, and procedures that researchers have used to build deep learning models that are effective in recommending refactoring solutions. Researchers have adopted various procedures related to data preprocessing, feature extraction, and data balancing to ensure that they build a deep learning model that is trained with high-quality data for accurate and efficient recommendations of refactoring solutions.

Table 4. Representative refactoring types.

| Level | Refactorings | References |
|---|---|---|
| Class | Extract class, Move class, Rename class | [3, 9, 65, 66, 69–71, 74] |
| Method | Extract method, Move method, Rename method | [4, 6, 9, 16, 17, 21, 44, 51, 53–55, 65, 66, 71, 71, 79, 98, 105] |
| Variable | Extract variable,Rename variable, Move attribute | [9, 59, 65, 66, 71] |

*5.3.1 Data Preprocessing.* According to our survey, researchers have employed various processes such as tokenization, lemmatization, stop word removal, noise removal, and normalization to ensure that their data is well-preprocessed and cleaned. Tokenization breaks texts into words, phrases, symbols, or other meaningful elements called tokens. This is used to split text into constituent sets of words. Lemmatization replaces the suffix of a word or removes it to obtain the basic word form. It is used for part of speech identification, sentence separation, and key phrase extraction. The goal of lemmatization is to group different inflected forms of a word so that they can be analyzed as a single item. Stop word removal involves filtering out common words that are considered to be of little value in understanding the meaning of a text. Noise removal refers to the process of reducing or eliminating irrelevant or unwanted information, often referred to as "noise," from a dataset. The goal of noise removal is to improve the quality of the data or signal for more accurate analysis or interpretation. For instance, Sagar et al. [79] had to remove and clean HTML tags since their data came from the web. Sagar et al. [79] also checked for special characters, numbers, and punctuation to remove any noise. Normalization refers to the process of transforming data into a standard scale. The goal is to bring the values of different variables or features into a comparable range, preventing one variable from dominating the analysis simply because of its larger scale. In this process, textual data may be converted to the standard required case, either lowercase, uppercase, camel case, etc. We also note through our literature search that other researchers, for example, Alenezi et al. [3], employed the process of basic data cleaning by first deleting all unnecessary features from their dataset, then finalizing the process by deleting refactoring type features and replacing the summation of refactoring feature.

*5.3.2 Feature Extraction.* Various types of features can be extracted to enable a deep learning model to recommend appropriate refactoring solutions. These features can include semantic, structural, code metrics, commit messages, or documentation. Most studies extract features from source code, which includes different metrics depending on the refactoring solution they want to suggest. However, some researchers, such as Aniche et al. [9], used process and ownership metrics instead of merely employing source code features. For the process metrics, Aniche et al. [9] collected five different types of metrics: the number of commits, the sum of lines added and removed, the number of bug fixes, and the number of previous refactoring operations. They calculated the number of bug fixes by using a heuristic whenever any of the keywords "bug, error, mistake, fault, wrong, fail, and fix" appeared in the commit message, and counted one or more bug fixes to that class. The number of previous refactoring operations was calculated based on the refactorings they gathered from the refactoring mining tool. For the code ownership metrics, Aniche et al. [9] adopted the suite ownership metrics as proposed by Bird et al. [11]. The number of authors was the total number of developers who had contributed to the given software artifact. The minor authors were the number of contributors who had authored less than 5% (in terms of commits) of an artifact. The major authors were the number of developers who contributed at least 5% to an artifact. With this, ownership was calculated as the proportion of commits achieved by the most active developer.

*5.3.3 Feature Embedding.* The primary studies have employed various tools and techniques for feature embedding. Embedding is a type of vectorization that represents words as dense vectors in a continuous space which captures semantic relationships between them. For instance, Cui et al. [16, 17] used code and graph embedding techniques to generate corresponding structural and semantic representations. They also used them to create hybrid representations. For code embedding, they used Code2vec [8] and Code2Seq [7]. Code2Vec is a neural network that automatically generates vectors from source code, while Code2Seq is a neural network that produces sequences from code snippets. For graph embedding techniques, they explored the use of Deepwalk [72], Node2Vec [30], Walklets [73], GraRep [13], Line [88], ProNE [106], and SDNE [97]. DeepWalk and Node2Vec employ random walk to construct sample neighborhoods for nodes in a graph based on a Skip-gram Natural Language Processing (NLP) model. The goal of Skip-gram is to maximize the likelihood of words appearing in a sliding window co-occurring. Walklets is another random walk-based graph embedding technique that explicitly encodes multi-scale relationships between nodes to produce multi-scale representations for them. GraRep is a matrix factorization-based graph embedding technique that constructs matrices from connections between nodes and factorizes them to produce the embedding result. Line calculates graph embedding results by specifying two functions, one for the first-order node proximity and the other for the second-order node proximity. ProNE is a fast and scalable graph embedding technique that was recently introduced. It includes two steps, the first is to effectively initialize graph embedding results by phrasing the problem as sparse matrix factorization, motivated by the long-tailed distribution of most graphs and their sparsity. The second stage is to propagate the initial embedding result using the higher-order Cheeger's inequality [46], aiming at capturing the graph's localized clustering information. SDNE employs deep autoencoders to generate embedding results. Other techniques also emerged from the researchers in the primary studies, for example, the employment of word embedding technologies, e.g., Word2Vec [51, 53, 79, 98], vector space models [65, 66], and CodeT5 [105] to achieve embedding.

*5.3.4 Data Balancing.* Researchers often use various techniques to address the issue of data imbalance for recommending refactoring solutions. These techniques include the Synthetic Oversampling Technique (SMOTE) and its variants (BLSMOTE, SVSMOTE, GraphSMOTE, etc.), UpSample, RUSBoost, Down sampling, and Random sampling. SMOTE technique is based on the oversampling approach in which synthetic examples are used for oversampling the minority class rather than oversampling with replacement. UpSample is used to improve the number of samples of the minority class by inserting zeros between the samples. RUSBoost is a hybrid approach of data sampling and boosting algorithm used to improve the performance of models trained on skewed data. To reduce the bias that may arise due to the use of imbalanced datasets, data balancing techniques are usually employed to create a balanced dataset. Most researchers employ the use of sampling techniques to achieve data balance in their datasets. In our survey, 82.00% of primary studies were found to employ sampling techniques in their variant forms. However, Pinheiro et al. [74] concluded that using balancing techniques that increase or decrease samples randomly is not the best strategy for improving datasets composed of code metrics.

### 5.4 Evaluation

*5.4.1 Datasets.* Different types of datasets are utilized to suggest accurate refactoring solutions through deep learning models. These datasets are carefully chosen to ensure that the deep learning model receives the appropriate data for making the right recommendations. Based on our survey, we found that researchers typically clone or download the subject projects from repositories and extract data relevant to their study refactorings. RefactoringMiner [93] was found to be the most popular tool for refactoring data mining tasks. RefactoringMiner was used by at least 91.30% of the primary studies. As shown in Figure 7, 86.96% of the researchers used publicly available projects, while only 13.04% used private projects with restricted access. Interestingly, 100% of the studies in this category the applications that were used for the experiments and evaluation were developed in the Java language. We also came across the work of Mastropaolo et al. [59], who created three datasets to train and evaluate their deep learning model. They built a large-scale dataset for training the model, tuning parameters, and performing an initial assessment of performance. Additionally, they created reviewed and developers datasets to further evaluate the performance of their technique.

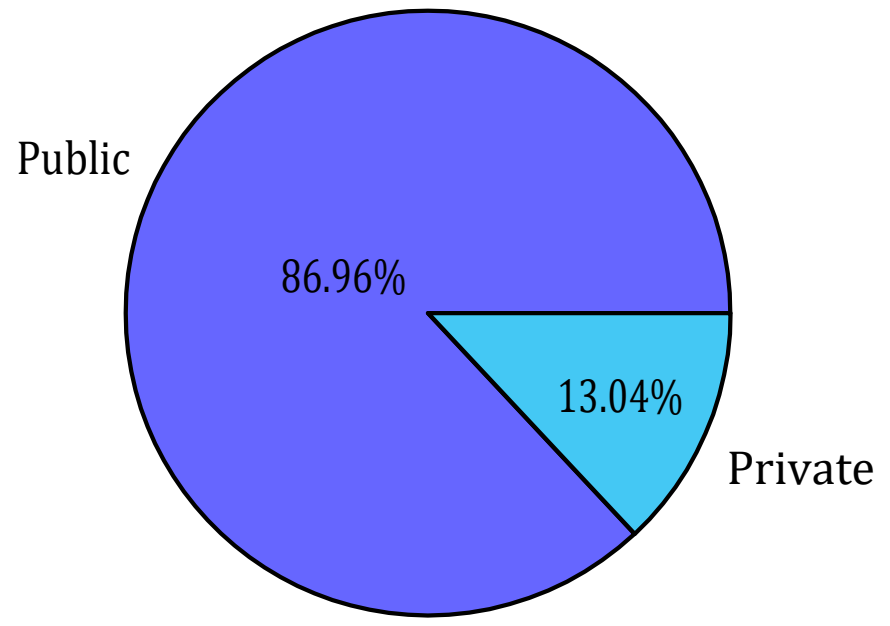

Fig. 7. Datasets

5.4.2 *Result Metrics.* As previously discussed, result metrics are used to measure the effectiveness of a particular approach. The standard metrics for this calculation are precision, recall, F-measure, and accuracy. Our literature search findings on the metrics used by researchers in this category are shown in Figure 8. According to Figure 8, F-measure, recall, and precision are the most commonly used metrics for evaluating various approaches in recommending refactoring solutions. F-measure was used in at least 95.65% of the primary works, while precision and recall were employed in 95.65% and 86.95% of the studies, respectively.

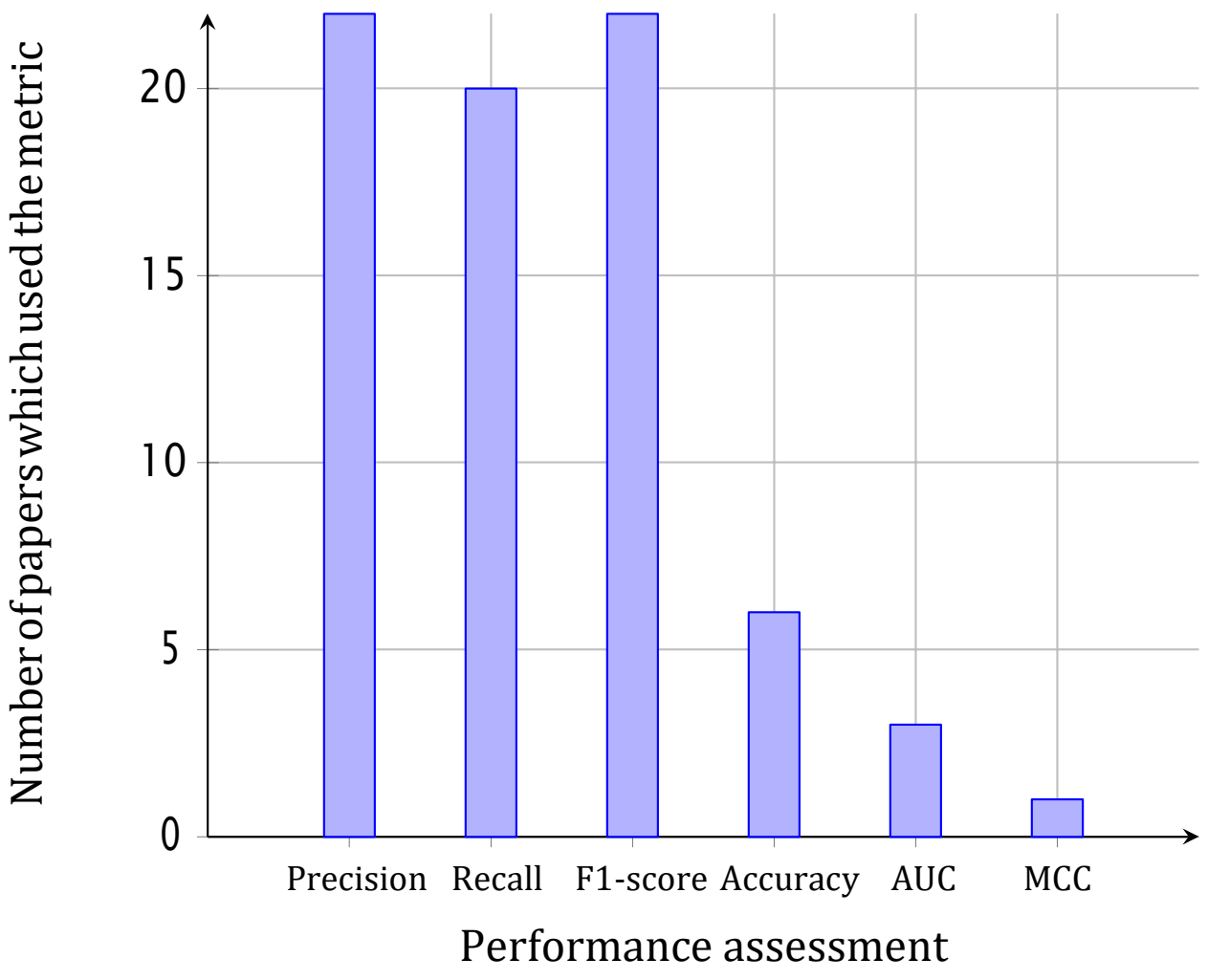

Fig. 8. Metrics used for performance assessment.

## 6 END-TO-END CODE TRANSFORMATION AS REFACTORING

The end-to-end code transformation as refactoring refers to the actual process of modifying an existing codebase to improve its structure, readability, maintainability, or performance without changing its external behavior. The end-to-end code transformation of refactorings differs from the other refactoring tasks (i.e., the detection of code smells, the recommendation of refactoring solutions) in that it involves the application of the suggested refactoring to the codebase, reviewing the refactoring code to ensure it adheres to coding standards, validating that the external behavior of the codebase remains unchanged, and updating any documentation to reflect the changes made during the refactoring, such as comments, inline documentation, and external documentation if necessary. This process focuses on actively making changes to the codebase which can include the actual renaming of variables, the moving of methods, extracting methods, simplifying complex expressions, restructuring code, and so on. In this section, we will explore how researchers have utilized deep learning models to conduct end-to-end code transformations as refactorings. Our specific focus will be on the technologies utilized for conducting the refactoring code transformations.

Szalontai et al. [87] have developed a method using deep learning to refactor source code, which was initially developed for the general-purpose programming language and runtime environment, Erlang. This approach has two main components: a localizer and a refactoring component. Together, they enable the localization and refactoring of non-idiomatic code patterns into their idiomatic counterparts. The method processes the source code as a sequence of tokens, making it capable of transforming even incomplete or non-compilable code. To do this, the source code is transformed into a sequence of tokens, using the Erlang module *tok* to tokenize the source code. The module obtains token types such as atom, integer, variable, etc. These tokens are then provided as input into the neural network to localize non-idiomatic functions. The neural network consists of convolutional, recurrent, and feedforward components. The tokens provided as input are embedded into a 64-dimensional vector space, and then a one-dimensional convolution is applied to each code chunk using 126=8 filters and a kernel size of 5. Two pooling operators are applied to the convolutional outputs, average and minimax. These two operations yield two intermediate representations for each chunk of the source code. The idiomatic alternative is generated using a recurrent sequence-to-sequence architecture with an attention mechanism. The non-idiomatic tokenized code is first fed to the encoder, which produces a hidden representation of the input sequence. This is achieved through the use of a Recurrent Neural Network consisting of four BiLSTM layers with 64 units each. The decoder uses a single LSTM layer with 256 units to generate an output sequence element by element, producing the idiomatic alternative. Both the localizer and the refactoring models were evaluated on a test set that was separated from the training data before the training process. The accuracy of the localizer was measured as the ratio of classified code chunks to the total number of chunks in the test set and was found to be 99.09%. For the refactoring component, the ratio of error-free transformations against the total number of attempted transformations was measured resulting in an accuracy of 99.46%. These results indicate that the presented models were trained successfully and are capable of performing refactorings that are similar to the ones in the training datasets.

Tufano et al. [96] quantitatively investigated the ability of a Neural Machine Translation (NMT) model to learn how to automatically apply code changes implemented by developers during pull requests. They harnessed NMT to automatically translate a code component from its state before the implementation of the pull request (pre-PR) and after the pull request has been merged (post-PR), thereby emulating the code changes that would be implemented by developers in the pull request. This is the first work that used deep learning techniques to learn and create a taxonomy from a variety of code transformations taken from the developer's pull requests. In this investigation, they first mined a dataset of complete and meaningful code changes performed by developers in merged pull requests, extracted from three Gerrit repositories (*Android*, *Google*, and *Ovirt*). Then they trained the NMT models to translate pre-PR code into post-PR code, effectively learning code transformations as performed by developers. *RNN Encoder-Decoder* and *Beam Search Decoding* were used as NMT models for this approach. The *RNN Encoder-Decoder* architecture was coupled with an attention mechanism which is commonly adopted in NMT tasks. The *RNN encoder* was used for encoding a sequence of tokens $x$ into vector representation while the *RNN Decoder* was used for decoding the representation into another sequence of tokens $y$. The primary purpose of the employed beam search decoder was to improve the quality of the generated token sequences by exploring multiple possible paths instead of simply selecting the most likely next token at each step. The NMT model was able to predict and learn from some transformations. This was used to develop a taxonomy of the transformations with three subcategories grouping the code transformation into *bug fixing*, *refactoring*, and *other*. The *refactoring* subtree included all code transformations that modified the internal structure of the system by improving one or more of its non-functional attributes without changing the system's external behavior. Under this subtree five subcategories were formulated, namely, *inheritance* (forbid method overriding by adding the final keyword to the method declaration, invoke overriding method instead of overridden by removing the super keyword to the method invocation and making a method abstract through the abstract keyword and deleting the method body), *methods interaction* (add parameter refactoring (i.e., a value previously computed in the method body is now passed as parameter to it), and broadening the return type of a method by using the Java wildcard (?) symbol), *readability* (braces added to if statements with the only goal of clearly delimiting their scope, the merging of two statements defining and initializing a variable into a single statement, the addition/removal of the this qualifier, to match the project's coding standards, reducing the verbosity of a generic declaration by using the Java diamond operator, refactoring anonymous classes implementing one method to lambda expressions, to make the code more readable, simplifying Boolean expressions, and merging two catch blocks capturing different exceptions into one catch block capturing both exceptions using the or operator), *naming* (renaming of methods, parameters, and variables), and *encapsulation* (modifying the access modifiers, e.g., changing a public method to a private one). The results showed that NMT models are capable of learning code changes and perfectly predict code transformations in up to 21% of the cases when only a single translation is generated and up to 32% when 10 possible guesses are generated. These results highlight the ability of the models to learn from a heterogeneous set of pull requests belonging to different datasets, indicating the possibility of transfer learning access projects and domains.

To facilitate the rename refactoring process and reduce the cognitive load of developers, Liu et al. [52] proposed a two-stage pre-trained framework called RefBERT. This framework is based on the BERT architecture and was designed to automatically suggest a meaningful variable name, which is considered a challenging task. The researchers focused on refactoring variable names, which is more complex than refactoring other types of identifiers, such as method names and type names. RefBERT uses 12 RoBERTa layers, which are a replicated version of the original BERT model with improved performance. The approach is based on three observations. First, rename refactoring is similar to Masked Language Modelling (MLM), a pretext task commonly used in pre-training BERT. MLM fills the masked part of a text according to its context. Similarly, rename refactoring aims to suggest a meaningful variable name according to the context. Therefore, MLM can be adopted for training an automatic rename refactoring model. Second, unlike the variable name prediction task, where only the context of the target variable is known, in rename refactoring, both the context of the target variable and the variable name before refactoring are known. Contrastive learning, which contrasts positive and negative samples for improving representation learning, is an ideal learning paradigm for automatic rename refactoring. The researchers expected the generated name to be close to the variable name after refactoring but far away from the variable name before refactoring. Third, unlike natural language text where words should follow a strict order to ensure grammatical correctness, subtokens in a variable name do not have such a restriction. Different orders of subtokens for a variable name do not significantly affect our understanding of the variable. Thus, the standard cross-entropy loss that emphasizes the strict alignment between the prediction and the target is suboptimal for automatic rename refactoring. RefBERT was trained to generate refactorings in two steps: Length Prediction (LP), where it predicts the number of tokens in the refactored variable name, and Token Generation (TG), where given the predicted number of tokens, RefBERT generates tokens in the refactored variable name. To train RefBERT, the researchers used the CodeSearchnet [40] and Java-Small [7] datasets in the pretraining stage. During the fine-tuning stage, they also used JavaRef and TL-Codesum [38] datasets. JavaRef was constructed by the researchers by applying data collection and preprocessing procedures on open-source datasets collected from GitHub. The experimental results demonstrated the effectiveness of RefBERT in automatic rename refactoring.

From the literature presented, we note from Table 5, that Recurrent Neural Networks (RNN) through its variants (LSTM and GRU) have mostly been used for the end-to-end transformation as refactorings. The usage of RNN was found in at least 75% of the studies. This was followed by Transformer technologies (e.g., BERT), which were utilized in 25% of the studies. Notably, to enhance the performance of the proposed techniques the researchers adopted the inclusion of other techniques in their approaches. Szantotai et al. [87] and Tufano et al. [96] used an attention mechanism to improve the model's ability to focus on relevant parts of the input sequence when generating the output sequence. In contrast, Liu et al. [52] used contrastive learning to contrast positive and negative samples for improving representation learning, and was used for automatic rename refactoring. The inclusion of contrastive learning helped the model to understand the context of the target variable and the variable name before refactoring.

From the presented studies, we note that researchers have employed various deep-learning techniques to perform the code transformation for different types of refactorings. Szalontai et al. focused on using a deep learning model to refactor nonidiomatic code patterns into idiomatic ones across various levels of code organization such as class, method, and variable. Typically, the choice of the appropriate level depends on the specific issues identified in the codebase, with refactorings like extract class, extract method, and rename variable being associated with these patterns. Szalontai et al.'s approach primarily targeted the general-purpose programming language Erlang. Conversely, Liu et al.'s [52] and Tufano et al. [96] approaches specifically targeted the renaming refactoring for variables and refactoring of non-function attributes, respectively, using the Java language. Thus, from Table 5, we observe that researchers are employing a specific approach to the conduction of end-to-end transformation of refactorings as found in 66.67% of the studies i.e., variable renaming and non-function attribute refactoring. In contrast, 33.33% of the studies opted for a general approach in the conduction of the refactoring transformation where the changes could be used at different levels (i.e., class, method, variable).

Table 5. Technologies used for the end-to-end refactorings transformation

| Technologies | Refactorings | Deployment tools/plaforms | References |
|---|---|---|---|
| LSTM+GRU+attention mechanism | Non idiomatic components | Erlang | [87] |
| NMT(RNN+Beam search)+attention mechanism | Non functional attributes | Java | [96] |
| BERT(12RoBERTA)+contrastive learning | Rename refactoring | RefBERT-Java | [52] |

## 7 MINING OF REFACTORINGS

Mining of refactorings refers to the automatic process of identifying and extracting instances of refactorings from existing codebases. To conduct the mining of refactorings for deep learning-based refactoring, traditional refactoring miners are used. The refactoring miners utilize various techniques such as static analysis, pattern recognition, and heuristic-based methods to identify and discover refactoring activities that were carried out within codebases. The outputs generated by these miners might consist of labeled examples that indicate where and how refactorings have been applied. These labeled examples serve as ground truth data, forming the foundation for training and evaluating deep learning models. By leveraging the outputs of traditional refactoring miners, large datasets of labeled refactorings, enabling the development of accurate and robust deep-learning models capable of automating software refactoring processes can be created. This integration of mining approaches with advanced deep-learning methodologies accelerates the advancement of intelligent tools aimed at enhancing code quality and maintainability.

Several traditional refactoring miners have been proposed by researchers to aid in the mining of refactorings. Tsantalis proposed RefatoringMiner [93] which represents the implementation of software entities as abstract syntax trees (ASTs), and computes the similarity between two entities according to the name-based similarity and the statement-based similarity. With such similarities, RefactoringMiner maps entities between two successive versions and leverages a sequence of heuristics to discover software refactorings based on the mapping. RefactoringCrawler developed by Dig et al. [20], is an analysis tool that detects refactorings that happened between two versions of a component. The strength of the tool lies in the combination of a fast syntactic analysis to detect refactoring candidates, and a more expensive semantic analysis to refine these candidates. Silva et al. [85] proposed RefDiff which utilizes static analysis and code similarity to detect various refactorings. It begins by tokenizing the source code of the project. Each code element (such as classes, methods, and fields) is transformed into a bag of tokens. Ref-Finder proposed by Prete et al. [75] encodes code elements (e.g., classes, methods, and fields) and their relationships using logic predicates to detect the refactorings. Liu et al. [49] proposed ReMapper an automated iterative approach used to match software entities between two successive versions for the discovery of refactorings. ReMapper takes full advantage of the qualified names, the implementations, and the references of software entities. It leverages an iterative matching algorithm to handle the interdependence between entity matching and the computation of reference-based similarity. Researchers have utilized some of these traditional refactoring miners to mine refactorings to train deep learning models in the process of refactoring as follows.

Although deep learning technologies have not yet been employed to distinguish refactorings from other source code modifications as RefactoringMiner or Ref-Finder do, deep learning technologies have been successfully employed to identify refactoring-containing commits by analyzing their associated commit messages. For example, Marmolejos et al. [58] developed a framework that used text-mining, natural language preprocessing, and supervised machine learning techniques to automatically identify and classify refactoring activities in commit messages. The framework focused on detecting Self-Affirmed Refactorings (SAR), which are refactoring activities reported in commit messages. The approach used a binary classification method to overcome the limitations of the manual process proposed in previous studies. The framework had four main parts. The first part involved preparing the data and processing the content of the commit messages to remove unnecessary and irrelevant information, as well as normalize the data. In the second part, the data was converted into hash values, with each hash value representing one or more features in the commit messages. The third part involved filtering the features to select only the most important ones from the dataset. Finally, in the fourth part, machine learning algorithms were trained and tested based on the selected features. The resulting two-class classifier was able to operate over unlabelled texts. For this approach, the authors used four classifiers, including Bayes Point Machine, Logistic Regression, Boosted Decision Tree, and Average Perceptron, as well as one deep learning-based classifier, Neural Network. The dataset used in this approach contained 1,208,970 refactoring operations, extracted using RefactoringMiner [93] from 3,795 open-source Java projects. From this dataset, the authors extracted commit messages containing the required patterns to create their refactoring dataset. Since the employed machine learning techniques could not directly identify text, the authors converted the collected data into hashes. They used the feature hashing technique, also known as a hashing trick, to derive features. In this technique, various words with varying lengths were mapped to different features based on the hash value. To determine the relevance of each attribute in the dataset, Chi-Square (CHI) was used to give a score, while Fisher Score (FS) was used to select a subset of features and score the distance between them. The machine learning classifiers were trained using a stratified train-test split methodology, where 70% of the rows of the transformed dataset from the selected features were used for training and the remaining 30% were used to measure the error rate. The approach proved to be efficient, as the authors obtained substantial accuracy.

Alomar et al. [6] aimed to investigate whether different words and phrases used in refactoring commit messages are unique to different types of refactorings. To achieve this, they employed machine learning techniques to predict refactoring operation types based on the commit messages. The prediction of the refactoring operation was formulated as a multiclass classification problem, which relied on textual mining of commit messages to extract relevant features for each class. The researchers collected a dataset of refactorings from 800 projects, where each instance presented a commit message and a refactoring type. They identified six preferred method-level refactorings, including extract method, inline method, move method, pull-up method, push-down method, and rename method. To identify relevant features, they used the n-gram technique proposed by Manning and Hinrich [57]. Nine supervised machine learning algorithms were applied, and the results were compared against a keyword-based baseline approach used in Murphy et al. [62]. The results revealed that the predictive accuracy for rename method, extract method, and move method ranged from 63% to 93% in terms of F-measure. Nevertheless, the model encountered challenges in accurately discerning between Inline Method, Pull-up Method, and Push-down Method, with F-measure scores falling within the range of 42% to 45%. Additionally, it's noteworthy that the keyword-based approach exhibited significantly lower performance compared to the machine learning models.

## 8 CHALLENGES AND OPPORTUNITIES

As the use of deep learning models in the domain of software refactoring continues to grow, it becomes imperative to closely look at the challenges and opportunities linked to their adoption. Despite showcasing promising capabilities in aiding different tasks of the refactoring process, there are still some challenges associated with their application. This section explores the challenges confronted by deep learning models in supporting the process of software refactoring, concurrently shedding light on prospective opportunities for future work.

### 8.1 Challenges

While deep learning models have proven to be effective in supporting the process of software refactoring, their adoption into this field is not without challenges. From our survey, we note the following challenges.

- Limited generalization of deep learning techniques across diverse paradigms is a significant concern. Many studies have developed approaches concentrating on specific code smells (e.g., feature envy, brain class, brain method) within a particular language, such as Java. This specialization restricts the applicability of these models to different code smells or programming languages. Given that code smells can manifest differently in diverse contexts, a model trained on one set of smells may not exhibit robust performance on others. Moreover, code smells often coexist and exhibit interactions. For instance, a lengthy method may signal a broader design issue, like a god class. Approaches focused on individual smells in the detection of code smells might overlook these intricate interactions, resulting in incomplete or inaccurate outcomes. Notably, based on the compiled primary works, a substantial majority (at least 87.50%) employed datasets developed in a singular programming language, i.e., Java, posing a challenge for the generalization of these approaches to other programming languages.

- Challenges in creating generic classification and feature engineering. Developing a universal classifier for diverse refactoring processes has proven challenging. This challenge is particularly evident in the detection of code smells, where different types of code smells necessitate distinct features and characteristics for accurate identification. Employing a one-size-fits-all approach may lead to diminished precision and recall, especially in studies utilizing sequential modeling-based deep-learning approaches to support software refactoring. Additionally, extracting pertinent features from abstract syntax trees or sequences of statements presents difficulties. The model's effectiveness relies on the accurate capture of both semantic and structural features. Establishing suitable mechanisms for feature extraction is pivotal for the success of deep learning-based approaches.

- Concerns about data quality and representativeness are pivotal factors influencing the performance of deep learning models. Certain approaches utilized automatically generated labeled data, potentially lacking accurate representation of real-world scenarios. Incorporating real-world examples could enhance the effectiveness of deep learning models.

- Limited adoption. Developers may be resistant to adopting automated refactoring tools supported by deep learning due to concerns about the reliability and trustworthiness of the generated code changes. Notably, most of the current studies on the end-to-end code transformation for refactorings by deep learning models have used prototype tools and non-industrial datasets which do not reflect the actual tools and data pools used by programmers.
- Need for continuous learning. Software systems evolve as software systems undergo maintenance and updates. Models trained on a static dataset may become outdated and may not effectively support the process of software refactoring in such without continuous learning mechanisms.

### 8.2 Opportunities

Amidst these challenges presented in Section 8.1, there exist some opportunities for advancing the integration of deep learning models in software refactoring. Some of the opportunities, according to our survey, are as follows.

- According to the taxonomy presented, deep learning models have been used for various tasks, including detecting code smells, recommending refactoring solutions, conducting refactorings, and mining refactorings. Our survey of the literature shows that deep learning models are primarily employed for detecting code smells, accounting for at least 56.25% in this category. The recommendation of refactoring solutions accounts for 33.33%, end-to-end code transformation as refactoring for 6.25%, and mining of refactorings for 4.17%. However, we did not find any significant study on the use of deep learning for software refactoring quality assurance in our literature search. This revelation shows that there is an imbalance in how deep learning has been employed in supporting refactoring tasks and thus, presents an opportunity for future work. It is highly valuable to fill this gap and ensure that all the tasks of software refactoring are fully supported by deep learning models.
- According to the primary studies presented, various types of deep learning models have been used for software refactoring. Most of the studies focused on utilizing hybrid models that combine deep learning models with other techniques to improve model performance. The data shows that at least 58.69% of the primary studies used hybrid approaches, while the remaining 41.31% used generic deep learning models. The latter encompassed sequential modeling, explainable and feedback-centric approaches that fit into developers' workflow. Among the deep learning models, CNN was the most commonly used model, found in at least 43.48% of the primary works, followed by RNN at 34.78%, and its variants (such as GRU, LSTM, GRU, etc.). The other 21.74% of the studies employed other deep learning models, such as GCN, GNN, ResNet, MLP, etc. However, only a few studies, such as Sharma et al. [80, 81], explored the use of transfer learning techniques to enable deep learning models trained on one project to be effectively used for refactoring in different projects. There is a need to carry out more exploration of transfer learning approaches as they can result in better generalization and minimize the need for extensive project-specific training data. Notably, Himesh et al. [64] and Yin et al. [104] were a few of the researchers who explored the inclusion of feedback and developments of explainable deep learning models for refactoring. To improve the adoption of deep learning models in the software refactoring process by software developers, it is necessary to do more explorations in the inclusion of feedback and explainability in the deep learning models.

- Our literature search has revealed that deep learning models have been predominantly used for method-level refactorings. Among the primary studies we surveyed, 55.41% applied deep learning models for refactorings at the method level, followed by 30.45% for class level, 10.12% for variable level, and 4.02% for other types of refactorings. The most frequent use of deep learning techniques was for detecting and applying move method and extract method refactorings, both of which occur at the method level of the codebase. It is worth noting that most of the work on identifying refactoring opportunities by detecting code smells has been focused on method-level code smells, particularly the feature envy smell which often leads to the recommendation and suggestion of move method refactoring. This opens up room for more future work in employing deep learning models for refactorings occurring at other levels than just the method level.

- Based on the literature, it has been found that deep learning models are effective in supporting the process of software refactoring. These models have outperformed the existing approaches or tools used in refactoring by achieving an average F-measure of 76%, which is a significant improvement of 30% on average compared to the state-of-the-art approaches. However, there is still a need to develop dynamic and adaptive deep-learning models that can continuously learn and adjust to changes in coding standards, project goals, and evolving best practices. This could involve using reinforcement learning approaches or other adaptive learning strategies. Additionally, according to our survey, most of the primary studies did not use industrial datasets, which makes it difficult to generalize the findings of most of the techniques. To improve the effectiveness of deep learning-based techniques for refactoring, it is necessary to incorporate more real-world industrial data.

## 9 CONCLUSIONS

In this paper, we have presented a survey on deep learning-based software refactoring. Our focus was on the process of software refactoring and how it can be supported by deep learning models. We have categorized the studies based on the specific refactoring task that was supported by deep learning models. Our taxonomy has identified five main categories which include the detection of code smells, recommendation of refactoring solutions, end-to-end code transformation as refactoring, quality assurance for refactoring, and mining of refactorings. We have presented key aspects under these categories which have provided insight into the research direction in the deployment of deep learning models for software refactoring.